\documentclass[aip,jcp,amsmath,graphicx,dvips,color]{revtex4-1}
\usepackage{mathptmx} 
\usepackage{courier} 
\normalfont
\usepackage[T1]{fontenc}
\usepackage[flushleft]{threeparttable}
\usepackage{multirow}
\usepackage{dcolumn}
\usepackage{bm}
\usepackage{graphicx}
\usepackage[version=3]{mhchem}
\usepackage{braket}
\usepackage{color}
\usepackage{booktabs}
\usepackage{enumerate}
\usepackage{ulem}
\usepackage[font=normalsize]{caption}

\renewcommand{\phi}{\varphi}
\renewcommand{\epsilon}{\varepsilon}

\begin{document}

\title{Electron paramagnetic resonance g-tensors from state interaction spin-orbit coupling density matrix renormalization group}
\date{\today}
\author{Elvira R. Sayfutyarova}
\affiliation{Department of Chemistry, Princeton University, Princeton, NJ 08543, USA}
\affiliation{Division of Chemistry and Chemical Engineering, California Institute of Technology, CA 91125, USA}
\author{Garnet Kin-Lic Chan}
\affiliation{Division of Chemistry and Chemical Engineering, California Institute of Technology, CA 91125, USA}

\begin{abstract}
We present a state interaction spin-orbit coupling method to calculate electron paramagnetic resonance (EPR) $g$-tensors from 
 density matrix renormalization group wavefunctions. 
We apply the technique to compute $g$-tensors for the \ce{TiF3} and \ce{CuCl4^2-} complexes,
a [2Fe-2S] model of the active center of ferredoxins, and a \ce{Mn4CaO5} model of the S2 state
of the oxygen evolving complex.
These calculations raise the prospects of determining $g$-tensors in multireference calculations
with a large number of open shells.
\end{abstract}

\maketitle

\section{Introduction}
Electron paramagnetic resonance (EPR) is a central tool in the study
of open-shell electronic structure as found in many complex transition metal systems, such as in the active sites of 
metalloenzymes.
The primary quantity measured is the transition energy between unpaired electron states
split by the external magnetic field. The proportionality between the effective magnetic moment
of the unpaired electron and its spin, namely the $g$-tensor, reports on the electronic environment of the electron. 
The utility of EPR arises from the high sensitivity of the $g$-tensor,
which thus yields invaluable information on the nuclear geometry and electronic structure.

The most common route to compute EPR $g$-tensors is through density functional theory (DFT)~\cite{vLenthe,ziegler1,ziegler2,Malkina,Kaupp,Neese01_2,g_DKH,g_DKH2,Komorovsky,Repisky,Rinkevicius}.
\textit{Ab initio} wavefunction methods to compute $g$-tensors have also been explored, using multireference configuration interaction
\cite{Lushington1,Bundgen,Lushington2,Bruna,Neese03,Brownridge,Neese07,Neese01,gSOCI},
the complete active space self-consistent field \cite{Vahtras,Bolvin,Ganyushin, Lan},
multireference perturbation theory \cite{Neese03,Bolvin,Vancoillie} 
and coupled-cluster theory \cite{Gauss, Bolvin}. A drawback of the electronic structure methods
listed above is that they are severely limited in terms of the number of open shells they can reliably handle,
restricting the kinds of transition metal complexes that can be studied. 
Recently, in Ref.~\cite{roemelt}, Roemelt presented an approach to compute $g$-tensors using a
  density matrix renormalization group (DMRG) description of the electronic structure, which is capable of
treating a significantly larger number of open shells than other techniques.
Here, we describe a related DMRG implementation to obtain $g$-tensors based  
on a state interaction spin-orbit coupling DMRG formalism~\cite{roemelt, elviras, reiher}. The main
methodological difference between our approach and that of Roemelt that we use is a more flexible representation of the interacting states.
Both approaches include spin-orbit coupling in the determination of the zeroth order wavefunction (i.e. they are first order approaches)~\cite{Bolvin,Ganyushin,Chibotaru,Lan},
rather than treating it together with the magnetic field (second order approaches)~\cite{Malkina,Kaupp,Lushington1,Bundgen,Lushington2,Bruna,Neese03,Brownridge,Neese07,Vancoillie,ziegler1,ziegler2,Vahtras,Rinkevicius}.
Together with Ref.~\cite{roemelt}, our work is a step towards obtaining $g$-tensors in transition metal complexes including a
rigorous treatment of a large number of open shells. We first describe the formalism in section \ref{sec:theory},
then proceed to demonstrate the calculation of $g$-tensors at the DMRG level for the \ce{TiF_3} and \ce{CuCl4^2-} complexes,
a [2Fe-2S] model complex, and a \ce{Mn4CaO5} model of the S2 state of the oxygen evolving complex (OEC).

\section{Theory}

\label{sec:theory}

We first recapitulate how to extract $g$-tensors from the spin and ab initio electronic Hamiltonians
in sections \ref{sec:theory1} and \ref{sec:theory2}. Other useful presentations that we draw on can be found
in Refs.~\cite{Bolvin,Vancoillie,Lan,Chibotaru}. In Section \ref{sec:theory3} we summarize how to treat spin-orbit coupling within the spin-orbit mean-field approximation.
Section \ref{sec:theory3} also describes how to calculate all the required quantities with DMRG wavefunctions and a state-interaction formalism,
as used in our earlier work in Ref.~\cite{elviras}.

\subsection{The spin Hamiltonian and ab initio Hamiltonian}
\label{sec:theory1}

The relationship between theory and actual EPR spectra is provided by the effective spin Hamiltonian.
The $g$-tensor arises as a parameter of the effective spin Hamiltonian
and it describes the Zeeman interaction between an external magnetic field $\mathbf{B}$ and an
effective spin $\mathbf{S}_\text{eff}$ of the molecule.
Other parameters of the effective spin Hamiltonian, the zero-field splitting tensor $\mathbf{D}$ and the hyperfine coupling tensor $\mathbf{A}$, 
define the splitting of energy levels in the absence of an external magnetic field .
 
Experimentally, the parameters $\mathbf{g}$, $\mathbf{D}$, $\mathbf{A}$ can all be extracted from EPR spectra. 
To derive them from quantum chemistry calculations one should, first, assume 
one-to-one correspondence between the spin-multiplet of an effective spin Hamiltonian 
and ab initio many-electron wavefunctions, describing actual electronic spin states of the paramagnetic complex.
The effective spin Hamiltonian in a magnetic field $\mathbf{B}$ is 
\begin{align}
  \hat{H}_\mathrm{spin} &= \mu_B \mathbf{B} \cdot \mathbf{g} \cdot \hat{\mathbf{S}}_\text{eff}+\hat{\mathbf{S}}_\text{eff}\cdot \mathbf{D}\cdot \hat{\mathbf{S}}_\text{eff}+\sum_R \hat{\mathbf{S}}_\text{eff}\cdot \mathbf{A}\cdot\hat{\mathbf{I}}_R , \label{eq:Hspin}
\end{align}
where  $\mu_B$ is the Bohr magneton,  $\hat{\mathbf{S}}_\text{eff}$  is the effective  spin operator, and $\hat{\mathbf{I}}_R$  is the
$R$th nuclear spin operator.
The parameters  $\mathbf{g}$, $\mathbf{D}$, $\mathbf{A}$ can be obtained by fitting the spectrum of Eq.~\ref{eq:Hspin} to the
theoretical spectrum of an ab initio electronic Hamiltonian $\hat{H}$ in a magnetic field, which can be expressed as:

\begin{align}
  \hat{H}&= \hat{H}_{0} +\hat{H}_\mathrm{Ze},
\end{align}
where $\hat{H}_{0}$ is the electronic Hamiltonian in the absence of the field and 
$\hat{H}_\mathrm{Ze}$ (the Zeeman interaction) groups together all terms with
an explicit dependence on the field. The Zeeman interaction can be expressed as
\begin{align}
  \hat{H}_{Ze} &= \mu_B(\hat{\mathbf{L}}+g_e \hat{\mathbf{S}}) \cdot \mathbf{B} \label{eq:zeeman}
\end{align}
where $\hat{\mathbf{L}}$ and $\hat{\mathbf{S}}$  are the total orbital and spin angular momentum operators,
and $g_e\approx 2.002319$ is the Land{\'e} factor for a free electron.

The electronic Hamiltonian $\hat{H}_{0}$ incorporates both spin-independent and spin-dependent relativitistic terms. 
The latter include the spin-orbit coupling (SOC), spin-spin coupling and hyperfine interactions.
In this work, we determine $\mathbf{g}$ from an ab initio calculation.
The $g$-tensor is predominantly determined by the spin-orbit coupling, which changes the effective spin of the unpaired electrons.
Therefore we omit hyperfine interaction and spin-spin coupling (required for $\mathbf{D}$ and $\mathbf{A}$) 
in the expression for $\hat{H}_{0}$ and write it in the form of:
\begin{align}
\hat{H}_{0}  = \hat{H}_\mathrm{SR} + \hat{H}_\mathrm{SOC}. \label{eq:H0}
\end{align}
We discuss the treatment of spin-orbit coupling further in Sec.~\ref{sec:theory3}.

\subsection{Extracting the g-tensor}
\label{sec:theory2}

We next consider how to extract the $g$-tensor by relating the contributions of the Zeeman interaction (\ref{eq:zeeman}) in
the ab initio Hamiltonian to that of  $\mu_B \mathbf{B} \cdot \mathbf{g} \cdot \hat{\mathbf{S}}_\text{eff}$ in the spin Hamiltonian (\ref{eq:Hspin}).
 We follow the reasoning of Gerloch and McMeeking, first established within ligand field theory~\cite{gerloch}
and later generalized to ab initio Hamiltonians~\cite{Chibotaru2,Bolvin,Vancoillie}. 
According to Kramers' theorem, in a molecular system with an odd number of electrons, i.e. with half-integer total spin, all states are at least
twofold degenerate in the absence of an external magnetic field; 
such pairs of degenerate states are referred to as Kramers pairs, and are
related by time-reversal symmetry. The ground-state is then described by (at least one) Kramers pair $(\Phi, \bar{\Phi})$
at zero magnetic field. The first-order effect of an external magnetic field is
to split this degenerate Kramers pair. The corresponding first order energy can be computed
from the Zeeman interaction in the ab initio Hamiltonian as well as with the spin Hamiltonian, which gives the expression for the $g$-tensor.

Using degenerate first order perturbation theory, the splitting of the eigenvalue pair $(\Phi, \bar{\Phi})$ is obtained by diagonalizing the first order interaction. 
Defining the symmetric quantity $G_{kl} =\sum_{n} g_{kn} g_{ln}$, or $\mathbf{G}=\mathbf{gg}^T$, one can then use the formula, proposed by Gerloch and McMeeking:
\begin{align}
   \mathrm{G}_{kl} &=2  \sum_{u,v=\Phi, \bar{\Phi}} \langle u|\hat{L}_k+g_e\hat{S}_k|v \rangle \langle v|\hat{L}_l+g_e\hat{S}_l |u \rangle \label{eq:ggt},
\end{align}
Finally, diagonalizing $\mathbf{G}$  yields the principal axes for the $g$-tensor, and the positive square roots of the three eigenvalues are the $g$-factors $g_1, g_2, g_3$.

The primary ab initio task is thus to obtain the matrix elements of $\hat{L}_k$ and $\hat{S}_k$ between
the degenerate Kramers ground-state pair in Eq.~\ref{eq:ggt}. This requires
determining the eigenvalues of the electronic Hamiltonian $\hat{H}_{0}$, including the spin-orbit coupling,
and computing the specific matrix elements in Eq.~\ref{eq:ggt}, as discussed in the next section.

\subsection{Spin-orbit coupling and matrix elements with DMRG}
\label{sec:theory3}

We determine spin-orbit coupled wavefunctions using the state interaction spin-orbit coupled DMRG (DMRG-SISO) that we described in Ref.~\cite{elviras}.
In this approach, the spin orbit operator $\hat{H}_\text{SO}$ is evaluated in a basis of  spin-adapted DMRG wavefunctions $\{ |\Psi_{I, S, M_S}\rangle \}$.

It is worth briefly mentioning the different choices of spin-adapted DMRG wavefunctions that can be used as the SISO basis.
For a set of $N$ orbitals (sites) the DMRG wavefunction amplitudes can be written in matrix product form. In
the so-called canonical form at site $i$, this is
\begin{align}
  |\Psi\rangle = \sum_{\{n\} } \mathbf{L}^{n_1} \ldots \mathbf{L}^{n_{i-1}} \mathbf{C}^{n_i} R^{n_{i+1}} \ldots \mathbf R^{n_N} \ket{n_1 \ldots n_N} \label{eq:mps}
\end{align}
where for a given occupation string, $\mathbf{L}^n$, $\mathbf{C}^n$, $\mathbf{R}^n$ are $M \times M$ matrices,
and the leftmost and rightmost boundary matrices are $1\times M$ row and $M \times 1$ column vectors respectively. 
There are different choices to optimize the matrices $\mathbf{L}^n$, $\mathbf{C}^n$, $\mathbf{R}^n$ in the DMRG sweeps.
In a so-called {\it state-averaged} DMRG calculation, a common renormalized basis (and thus a common set of $\mathbf{L}^n$, $\mathbf{R}^n$ matrices)
is optimized for all the electronic states, and only the $\mathbf{C}^n$ matrix is unique to each state.
The state-averaged DMRG representation was used in Ref.~\cite{roemelt} by Roemelt to represent all the states (including of different spin)
entering into the SISO procedure. An alternative, state-specific, approach is to have
 different sets of $\mathbf{L}^n$, $\mathbf{C}^n$, $\mathbf{R}^n$ for different states in the calculations. 
This was the approach in our previous work~\cite{elviras}, where we used a different set of $\mathbf{L}^n$, $\mathbf{C}^n$, $\mathbf{R}^n$ for states of different spin
(although we used a state-averaged representation for states of the same spin). For a given bond dimension $M$, the state-averaged approach 
reduces the amount of computation, but results in a lower accuracy for each individual eigenstate compared to the state-specific approach. Finally, in this work we also consider
a cheaper approach, where in each spin-sector, we use the $\mathbf{L}^n$, $\mathbf{R}^n$ {\it optimized only for the lowest state} and represent
the excited states by different $\mathbf{C}^n$ matrices, determined at the middle of the sweep. We denote this choice as ``ground-state specific'', and it has the advantage of avoiding
the large number of Davidson steps required to solve for multiple states when optimizing the $\mathbf{L}^n$, $\mathbf{R}^n$ tensors. This allows us to compute
a larger number of DMRG states to use for the SISO basis.

For the spin-orbit operator $\hat{H}_\text{SO}$, we use the spin-orbit mean-field (SOMF) approximation. This has been shown
to approximate the effects of the full one and two electron Breit-Pauli SOC operator very accurately~\cite{Marian:SOC,Hess:AMFI,Tatchen,Neese:SOMF}.
In second quantization the one-electron SOMF Hamiltonian can be written as~\cite{Neese:SOMF,Malmqvist:RASSI_SO}:
\begin{align}	
\hat{H}_\mathrm{SOMF} &= \sum_{ij}\left ({V}_{ij}^{x}\hat{T}_{ij}^{x} +{V}_{ij}^{y}\hat{T}_{ij}^{y}+{V}_{ij}^{z}\hat{T}_{ij}^{z} \right ) ,\label{eq:somf}
\end{align}
where 
 $\hat{T}_{ij}^{x,y,z}$ are the Cartesian triplet excitation operators \cite{helgaker:purplebook,MolProp}:
\begin{align}
\hat{T}_{ij}^{x}&=\frac{1}{2}\left ( a_{i\alpha}^{\dagger}a_{j\beta} + a_{i\beta}^{\dagger}a_{j\alpha}\right )  
\\ \hat{T}_{ij}^{y}&=\frac{1}{2i}\left ( a_{i\alpha}^{\dagger}a_{j\beta} - a_{i\beta}^{\dagger}a_{j\alpha}\right )
\\ \hat{T}_{ij}^{z}&=\frac{1}{2}\left ( a_{i\alpha}^{\dagger}a_{j\alpha} - a_{i\beta}^{\dagger}a_{j\beta}\right ).
\end{align}
and  $V_{ij}^{x,y,z}$ is an effective set of one-electron integrals, obtained as:
\begin{align}
\braket {i| \hat{\mathbf{V}}| j}&=\braket {i| \hat{\mathbf{h}}_{1} | j}+\sum_{kl}D_{kl}\left \{\braket {ik| \hat{\mathbf{g}}_{12} | jl} -\frac{3}{2}\braket{ ik| \hat{\mathbf{g}}_{12} | lj} -\frac{3}{2}\braket {ki| \hat{\mathbf{g}}_{12} | jl}  \right \} \label{eq:Vint}
\end{align}
where $D_{kl}$ is the single-particle (not necessarily idempotent) density matrix element that takes into account single occupancy
due to unpaired spins;
the one- and two-electron operators are
\begin{align}
  \hat{\mathbf{h}}_{i}&=\frac{\alpha ^{2}}{2}\sum_{A}Z_{A}r_{iA}^{-3}\hat{\mathbf{l}}_{iA},  \label{eq:bpterms1}
\\  \hat{\mathbf{g}}_{ij}&=-\frac{\alpha ^{2}}{2}\hat{\mathbf{l}}_{ij}r_{ij}^{-3}, \label{eq:bpterms2}
\end{align}
where $\alpha$ is the fine structure constant, $\hat{\mathbf{r}}_{i}$, $\hat{\mathbf{p}}_{i}$ are the position and momentum operators of the $i$th electron,
 $r_{ij}= |\hat{\mathbf{r}}_{i}-\hat{\mathbf{r}}_{j} |$,  $\hat{\mathbf{l}}_{ij} =\hat{\mathbf{r}}_{ij}\times\hat{\mathbf{p}}_{i}$ , 
 $\hat{\mathbf{l}}_{iA} =\hat{\mathbf{r}}_{iA}\times\hat{\mathbf{p}}_{i}$,  $r_{iA}= |\hat{\mathbf{r}}_{i}-\hat{\mathbf{R}}_{A}|$, and $Z_{A}$ denotes the nuclear charge of the $A$th nucleus.
Note that this form of the SOMF operator is only valid for doublet ground states;
for  $S>1/2$ there is an additional correction $\frac{1}{2} \sum_{mn}D_{mn} \braket{ im| \hat{\mathbf{g}}_{12} | jn}$, where $m,n$ are the singly occupied orbitals \cite{Waroquier}. However, we have not considered this correction here as all our ground states are doublets.
As we  work with a spin-adapted basis, internally we do not use the Cartesian triplet operators,
but rather the spherical tensor triplet operators. These are related to the Cartesian triplet operators through the linear transformation \cite{helgaker:purplebook}:
\begin{align}
\hat{T}_{ij}^{x}&=\frac{\hat{T}_{ij}^{1,-1}-\hat{T}_{ij}^{1,1}}{2} \\
\hat{T}_{ij}^{y}&=\frac{\hat{T}_{ij}^{1,-1}+\hat{T}_{ij}^{1,1}}{2i}  \\
\hat{T}_{ij}^{z}&=\frac{1}{\sqrt{2}} \hat{T}_{ij}^{1,0}.
\end{align}
Using this form of the spin-orbit operator,  we evaluate the Hamiltonian in Eq.~\ref{eq:H0} in the 
basis of spin-adapted DMRG states using the transition density matrix algorithm  described in Ref.~\cite{elviras}. Diagonalizing this yields the spin-coupled Kramers pairs wavefunctions and energies.

Additionally, to determine the $\mathbf{G}$ matrix and $g$-tensors following the procedure in the preceding section, 
we need the matrix representations of the operators $\hat{L}_x, \hat{L}_y, \hat{L}_z$ and $\hat{S}_x, \hat{S}_y, \hat{S}_z$ for the ground-state Kramers pair.
We can obtain these from the matrix elements in the basis of spin-adapted DMRG wavefunctions $\{ |\Psi_{I,SM_S}\rangle\}$,
\begin{align}
\langle\Psi_{I, S' M'_{S}}| \hat{S}_{k}  | \Psi_{J,SM_{S}}\rangle &= \sum_{p}\langle\Psi_{I, S' M'_{S}}| \hat{T}_{pp}^{k}  | \Psi_{J,SM_{S}}\rangle, \qquad   k=x,y,z   \label{eq:Sk} \\
\langle\Psi_{I, S' M'_{S}}| \hat{L}_{k}  | \Psi_{J,SM_{S}}\rangle &=\sum_{ij} \langle\Psi_{I, S' M'_{S}}|  \hat{T}_{ij}^{0,0}  | \Psi_{J,SM_{S}}\rangle \delta_{SS'} \delta_{M_SM'_{S}}L_{ij} \label{eq:Lk}  
\end{align}
where $ \hat{T}_{ij}^{0,0}=\frac{1}{\sqrt{2}}\left ( a_{i\alpha}^{\dagger}a_{j\alpha} + a_{i\beta}^{\dagger}a_{j\beta}\right ) $ is the singlet operator, 
and $L_{ij} =\langle i |(\hat{\mathbf{r}}-\hat{\mathbf{R}}_{0})\times\hat{\mathbf{p}}|j \rangle$ is the orbital angular momentum
integral between the molecular orbitals $i$ and $j$, calculated with respect to an arbitrary gauge origin $\mathbf{R}_{0}$, chosen here to be 
the coordinates of the transition metal centers.
The singlet operator matrix element can be computed following Ref.~\cite{Sandeep:dmrg}. Once these matrix representations are obtained,
they are contracted with the expansions of the Kramers pairs in the spin-adapted basis $\{ |\Psi_{I,SM_S}\rangle\}$ to obtain representations in the Kramers basis.
\section{Results and Discussion}
\label{sec:results}
We implemented the above method as a stand-alone code and as a module within a development version of \textsc{PySCF}~\cite{pyscf}. 
Spin-orbit integrals in Eq.~\ref{eq:bpterms1}, \ref{eq:bpterms2} were computed using \textsc{PySCF}. Additional CASSCF calculations
were carried out using the \textsc{Molpro} package and geometry optimization of the [2Fe-2S] complex was carried out using \textsc{ORCA}.
 
All three components of the ${g}$-tensor (transformed to the principal axes, denoted arbitrarily as $x,y,z$) can be
expressed as shifts from the Land\'{e} $g$-factor
 \begin{align}
g_k=g_e + \Delta g_k, \qquad   k=x,y,z
\end{align}
In systems with axial symmetry $g_x = g_y = g_{\perp}$ and $g_z = g_{\parallel}$. 
For some systems, we present $\Delta g_k$ shifts instead of the full $g_k$-values. 

\subsection*{A. \ce{TiF3}}

We begin by considering the \ce{TiF3} complex. This has been widely used as a benchmark system for $g$-tensor calculations.
We used ANO-RCC basis sets contracted as [4s3p2d1f] for F and [7s6p5d3f2g1h] for Ti, and a 
 $D_{3h}$ symmetric complex with a Ti--F bond distance of 1.774{\AA}, following Ref. \cite{Vancoillie};
the calculations employed the  $C_{s}$ subgroup, with the 
Ti atom at the origin, one F atom on the y axis and other two in the xy plane. 

A minimal active space for this complex is formed by the $3d$ orbitals of Ti, giving a (1e,5o) active space. However,
there is no correlation within this space. To construct a larger active space,
we further included the $2s, 2p$ orbitals of each F atom, and the $3s, 3p, 4s, 4p$ orbitals of Ti. This gives a
(33e, 25o) active space that includes the dominant core-valence and valence-virtual correlation effects, both
for the energies and the density matrices.

The ground $^2A'_1$ state has one unpaired electron in the $3d_{z^2}$ orbital. The lowest excited states are 
metal centred ligand-field states; charge-transfer states have been seen to give negligible contributions to the $g$-values
in previous studies~\cite{Vancoillie}.
Thus we choose the lowest 5 ligand-field states for the state-interaction basis. We first optimized $3d$ orbitals using state-averaged CASSCF in the (1e,5o) active space 
using \textsc{Molpro}, with scalar relativistic effects included with the second-order Douglas-Kroll approximation~\cite{DKH1,DKH2,DKH3}. We then used DMRG in a larger (33e,25o) active space (using the (1e,5o) CASSCF orbitals).
The DMRG energy for each state was converged to better than $10^{-6} E_h$ with a bond-dimension of $M$ = 3000.

The 5 lowest electronic states, without SOC, are presented in Table \ref{tab:tab1}. The calculated and reference $g$-tensors from
the literature are summarized in Table \ref{tab:tab2}.
\begin{table}[h] 
\begin{threeparttable}
\caption{The electronic states of \ce{TiF3} (cm$^{-1}$) from this work, experiment, and previous theoretical studies.}
\label{tab:tab1}
\begin{tabular}{c|c| c|c|c|c|c|c }
\hline \hline
State                  &CASSCF         & DMRG& MRCISD+Q*\cite{solomonik}&CCSD(T)\cite{solomonik} &CCSD(T) \cite{Bolvin}   &CASPT2 \cite{Bolvin}  & CASPT2  \cite{Vancoillie}  \\
                          &(1e,5o)        &  (33e,25o)   & (1e, 5o)&  &               &(7e, 8o) & (17e,13o)          \\
\hline
$X ^2A_1^{'}$   &              &         &   	          &         &         &              &                    \\                      
$1 ^2E''         $   & 3502     & 5785  &4414 &5181  &  5600    &  3700     & 4789           \\
$1 ^2E'         $    & 20158  & 22538 &19379&19855&             &               &                  \\
\hline \hline
\end{tabular}
\end{threeparttable}
\end{table}


\begin{table}[h]
\begin{threeparttable}
\caption{ Calculated and experimental $\Delta g$ shifts for \ce{TiF3} (in ppt). }
\label{tab:tab2}
\begin{tabular}{ c|c|c|c|c|c|c|c|c|c}
\hline \hline
g-values         & DMRG-SISO&CASPT2\cite{Bolvin}&CCSD(T) \cite{Bolvin} &CASPT2 \cite{Vancoillie} &MRCI \cite{Brownridge}& SORCI \cite{Neese03} &ZORA\cite{vLenthe}&BP86  & Exp.\cite{deVore}\\
                              & (33e,25o)           &(7e, 8o)   &                                  &(17e,13o)                       &                                          \\
\hline
$\Delta g_{\perp}$          &  -113.7  &   -125.3  & -118 & -143.5 (I)   & -115.3    &-75.5        &-79.7  &    -30.7\cite{Neese01} &-111.3$^a$   \\
                                      & &                &       &-147.6  (II)  &              &                 &          & -36.0\cite{Kaupp}      & -123.7$^b$\\
                                      &    &                &       &                  &               &                 &          &-26.6\cite{Malkina}         &             \\
$\Delta g_{\parallel} $  &-1.6 & -2.4        &  -1.6 &   0.0 (I) & -0.9         &-0.1          &-1.1   &    -0.9 \cite{Neese01}        & -11.1$^a$    \\
                            &    &               &         & -2.5 (II) &               &                &          &-1.3\cite{Kaupp}              & -3.7$^b$ 	\\
                            &      &               &         &             &	              &                &          &-1.1 \cite{Malkina}         & \\
\hline \hline
\end{tabular}
\begin{tablenotes}
\item [a] From the EPR spectrum of \ce{TiF3} in solid neon at 4K.
\item [b] From the EPR spectrum of \ce{TiF3} in solid argon at 4K.
\end{tablenotes}
\end{threeparttable}
\end{table}

The (1e,5o) CASSCF calculation underestimates the energy of the lowest excited $1 ^2E''$ state compared to larger active space calculations.  The CASPT2 energies from Ref.\cite{Bolvin}, obtained with the (7e, 8o) active space (obtained from the (1e,5o) active space
by including three additional occupied ligand orbitals) confirms this observation.   
The DMRG (33e,25o) energies are in good agreement with the literature CCSD(T) energies. 
Including SOC removes the degeneracy of the $1 ^2E''$ and $1 ^2E'$ states and results in 144 cm$^{-1}$ and 228 cm$^{-1}$ splittings
at the DMRG level, respectively, giving two pairs of states with excitation energies 5719 cm$^{-1}$, 5863 cm$^{-1}$ and 22428 cm$^{-1}$, 22656 cm$^{-1}$.

For the $g$-tensor, the DFT-based approaches significantly underestimate $\Delta g_{\perp}$  \cite{Neese01, Kaupp,vLenthe,Malkina}.
The wavefunction-based $g$-shifts, including from the (33e,25o) DMRG-SISO calculation, are all in quite good agreement with experiment; there is
particularly close agreement between the DMRG-SISO $g$-values and those obtained from CCSD(T).

\subsection*{B. \ce{CuCl4^2-} }

We next consider the square planar \ce{CuCl4^2-} complex. This can be viewed as a model complex for copper sites in
blue copper proteins,
such as plastocyanin.  
We used ANO-RCC basis sets contracted to [5s4p2d1f] for Cl and [7s6p5d3f2g1h] for Cu,
and a $D_{4h}$ symmetric complex with a Cu--Cl distance of 2.291{\AA} as in Ref. \cite{Vancoillie}.
For the active space, we considered a Cu $3d, 4s, 4d$  (9e, 11o) active space, a minimal active
space including double-shell effects. We also considered two larger active spaces: one with additional
$3s$, $3p$, $4p$ Cu orbitals and four $\sigma$-orbitals formed by the $3d$ orbitals of Cu and the $3p$ orbitals of Cl atoms,
giving a (25e, 22o) active space; and one that further incorporates the $3p$ orbitals of the Cl atoms, that
provide $\pi$-interactions with the metal, giving a (41e, 30o) active space.

We first optimized the $3d$ orbitals using state-averaged CASSCF with the (9e,11o) active space using \textsc{Molpro}~\cite{molpro}. 
Scalar relativistic effects were included using the second-order Douglas-Kroll approximation~\cite{DKH1, DKH2,DKH3}. 
DMRG calculations were then performed with the (25e, 22o) and (41e,30o) active spaces for the lowest 5 ligand-field excited states.
These 5 states were used as the SISO basis. 
The DMRG energy for each state was converged to better than $10^{-6} E_h$ accuracy using a bond-dimension of $M$ = 3000.
The electronic states with and without SOC are presented in Table \ref{tab:tab3}. Calculated and reference g-tensors from the
literature are summarized in Table \ref{tab:tab4}.

\begin{table}[h]
\begin{threeparttable}
\caption{The electronic states of \ce{CuCl4^2-} (cm-1). The SOC-corrected energies from DMRG-SISO are given in parentheses. }
\label{tab:tab3}
\begin{tabular}{c|c|c|c|c|c|c|c  }
\hline \hline
 State & CASSCF& \multicolumn{2}{c|}{DMRG} & \multicolumn{2}{c|}{DMRG+NEVPT2} &   CASPT2  \cite{Vancoillie}          & Exp.* \\
          &(9e,11o)   &   (25e, 22o) &(41e,30o)&(25e,22o)&(41e,30o)  &(11e, 11o)        &            \\
\hline    
$1 ^2B_{2g}$     &6735   &  9438 &9079    &10780 & 10459         & 11321       & 10500 \cite{Willet}  , 12000\cite{Solomon1} \\
                           &          &(9382) &(9025)  &(10708)&(10392)              & &\\
$1 ^2E_{g}$       &8925    &  11572&11274 &12884 &12673 &13379      &12800 \cite{Willet} , 13500\cite{Solomon1} \\
                           &          &(11192,11480) &(11063,11179)  &(12465,12778)&(12490,12561)& &\\
$1 ^2A_{1g}$     &9918  & 12226 &12369&13476 &13886 &14597       &16500\cite{Solomon1} \\
                           &          &(13193) &(13164)  &(14460)& (14632)& &\\
\hline \hline
\end{tabular}
\begin{tablenotes}
\item [*] Polarized absorption spectrum for a single-crystal $D_{4h}$ \ce{[CuCl4]^2-}.
\end{tablenotes}
\end{threeparttable}
\end{table}

\begin{table}[h]
\begin{threeparttable}
\caption{Calculated and experimental $\Delta g_{\perp}$ and $\Delta g_{\parallel} $ shifts of \ce{CuCl4^2-} (ppt).  }.
\label{tab:tab4}
\begin{tabular}{ c|c|c|c|c|c|c|c  }
\hline \hline
 g-values      & \multicolumn{2}{c|}{DMRG-SISO} & \multicolumn{2}{c|}{DMRG-SISO+NEVPT2} &LFT*\cite{Solomon2}& CASPT2  \cite{Vancoillie}        & Exp.  \\
                                &(25e,22o)&(41e,30o)&(25e,22o)&(41e,30o)&     &(11e, 11o)                           &            \\
\hline
$\Delta g_{\perp}$               & 100.6 &82.6  &92.3&77.1&117 &  96.1 (I); 77.7 (II)                 &47\cite{Chow} , 38\cite{Solomon1}\\                      
$\Delta g_{\parallel} $          &  517.9 &529.8&458.7&464.3&531 & 466.0 (I);437.7 (II)                  & 230\cite{Chow}, 219\cite{Solomon1}\\
\hline \hline
\end{tabular}
\begin{tablenotes}
\item [*] These are obtained by fitting the ligand field excited state energies, obtained from polarized single crystal electronic absorption spectroscopy, to the $g$-tensor expression in the ligand field approximation.
\end{tablenotes}
\end{threeparttable}
\end{table}

The (9e,11o) active space significantly underestimates the excitation energies of all the states.
Including the near-valence orbitals of Cu and the $3p$ orbitals of Cl atoms in the DMRG calculation
recovers an important piece of the dynamic electron correlation, shifting the
excitation energies upwards by $\approx$ 2300-2700 cm$^{-1}$.
To verify the effects of dynamic correlation, we have also carried out DMRG-NEVPT2 calculations for each state~\cite{dmrg:nevpt2};
for the corresponding $g$-tensor calculations, the energies were used to shift the SISO matrix elements by $\Delta \hat{H}_\mathrm{SR}^{IJ}=0.5(\Delta E^I_{PT2}+\Delta E^J_{PT2}) \langle \Psi_I|\Psi_J\rangle$~\cite{Malmqvist:RASSI_SO}.
The DMRG-NEVPT2 excitation energies are shifted further upwards, giving improved agreement with the experimental 
excitation energies (see  Table \ref{tab:tab4}). Including the SOC in the DMRG-SISO has a large effect on the $1^2A_{1g}$ excitation,
although it remains below the experimental number.

The $g$-values calculated with the different theoretical methods are roughly comparable. In the DMRG-SISO
calculations,  the effect  of increasing the active space size or including dynamic correlation is to lower $\Delta g_{\perp}$
but raise $\Delta g_{\parallel}$. However,  the DMRG-SISO $g$-values remain too large when compared with experiment, almost by a factor of 2.   This is consistent with earlier CASPT2 calculations~\cite{Vancoillie} which also found an overestimation by a factor of 2.
In Ref.~\cite{Vancoillie} it is argued that as the CASPT2 excitation energies are quite accurate for this compound, and that the error
must arise in the density matrices, which yield too large matrix elements for the spin-orbit coupling operator due to
too much ionic character in the Cu--Cl bond. Our results indicate that this remains true even when the density matrices are relaxed
in the larger active space treated by DMRG.

\subsection*{C. \ce{[2Fe-2S]^+}}
We now consider the \ce{[Fe2S2(SCH3)4]^3-} complex. This can be considered to be a model of the active site in certain iron-sulfur proteins,
such as the ferredoxins in their reduced form.
To assess geometrical effects we performed calculations at three different geometries: the relaxed geometry from Ref. \cite{sharma:FeS} (geometry I), which was optimized by the authors at the def2-SVP/BP86 level of theory, and two geometries, which we optimized
at the def2-TZVP/TPSSh level of theory with (geometry II) and without inclusion of solvation effects (geometry III), using \textsc{ORCA}~\cite{orca}. Solvation was included via the COnductor-like Screening Model (COSMO) with a dielectric constant 4.0, which crudely imitates a protein environment. 
Table \ref{tab:tab5} summarizes the structural parameters for the three geometries used for the model \ce{[Fe2S2(SCH3)4]^3-} complex
as well as the geometries of two high-resolution X-ray structures of two reduced ferredoxin species: from the green alga \textit{Chlorella fusca} \cite{geom} and from the cyanobacterium \textit{Anabaena} PCC7119 \cite{geom2}. 
As one can see,  geometry II, obtained by including solvation effects, mimics the  ferredoxin active centre
better than the other model geometries.

\begin{table}[h]
\begin{threeparttable}
\caption{Structural parameters for the reduced \ce{[2Fe-2S]} cluster, obtained from optimized geometries of a model compound \ce{[Fe2S2(SCH3)4]^3-} and high-resolution X-ray crystal structures of different reduced ferredoxins.}
\label{tab:tab5}
\begin{tabular}{c|c|c|c|c|c   }
\hline \hline
Bond lengths& \multicolumn{3}{c|}{Optimized geometries}&\multicolumn{2}{c}{X-ray structures from} \\  
angles   & I  & II& III  & \textit{Chlorella fusca}\cite{geom}&\textit{Anabaena}PCC7119\cite{geom2}\\
\hline
Fe1-Fe2, {\AA}           & 2.914 & 2.827 &2.775&2.733&2.749\\
Fe1-S1, {\AA}            & 2.365 & 2.355 &2.326&2.230&2.293\\
Fe2-S1, {\AA}            & 2.267 & 2.214 &2.222&2.196&2.235\\
Fe1-S2, {\AA}            & 2.379 & 2.357 &2.336&2.224&2.261\\
Fe2-S2, {\AA}            & 2.260 & 2.178 &2.213&2.157&2.178\\
Fe1-S1-Fe2, $^{\circ}$  & 77.9 & 76.4&75.2 &76.3&74.7\\
Fe1-S2-Fe2, $^{\circ}$  & 77.8 & 76.3 &75.1&77.2&76.5\\
S1-Fe1-S2, $^{\circ}$  & 98.8 & 99.2  &101.1&101.4&101.8\\
S1-Fe2-S2, $^{\circ}$  &105.5 & 108.1&108.5&104.8&106.4\\
\hline \hline
\end{tabular}
\end{threeparttable}
\end{table}

To determine a suitable active space at each geometry, we first carried out an unrestricted Kohn-Sham
(UKS) BP86/TZP-DKH calculation of the high spin state with $S = 9/2$. Scalar relativistic effects were included using the exact-two-component (X2C) approach \cite{x2c} implemented in \textsc{PySCF}.
From the alpha and beta UKS orbitals, we constructed  unrestricted natural orbitals (UNOs).
From the UNO occupations, the orbitals were separated into three subspaces: doubly occupied, singly occupied and virtual molecular orbitals. 
Next, localized orbitals were constructed by projecting atomic orbitals
into these 3 spaces (e.g. a localized core 1s orbital is obtained by projecting a 1s orbital into the doubly occupied space)
followed by a subsequent orthonormalization within the spaces. By population analysis and visualization of the projected AO's 
we determined a suitable active space.
In this way we obtained a (31e,36o) active space including the (1) $3d$, $3d'$, $4s$ orbitals for Fe, 
(2) three $3p$ and two lowest-energy $3d$ orbitals on each bridging S atom, (3) an additional $3p$ orbital  on each ligand S atom.

\begin{table}[h]
\begin{threeparttable}
\caption{Dependence of $g$-values of the reduced \ce{[2Fe-2S]} complex on the number of electronic states included in the doublet and quartet manifolds (for geometry I). 
The DMRG energies were converged using $M$=3000.}
\label{tab:tab6}
\begin{tabular}{ m{2cm}m{2cm} m{2cm}m{2cm}	}
\hline \hline
  $g$-values   & \multicolumn{3}{c}{\# of doublet and quartet states in DMRG-SISO*} \\
                      &2 + 2&3 + 3&5 +5\\
\hline
$g_x$                & 1.989&1.888&1.807\\                     
$g_y $               & 2.004&1.989&1.931\\
$g_z$                & 2.006&2.004&1.969\\
\hline \hline
\end{tabular}
\end{threeparttable}
\end{table}

We carried out DMRG calculations for the doublet, quartet, and hextet states, i.e. with $S = 1/2, 3/2, 5/2$.
Note that sextet states and higher do not directly spin orbit couple with the ground doublet state,
however, they can contribute indirectly to the $g$-tensor via coupling with lower spin states, changing their energies.
Sharma et al~\cite{sharma:FeS} have shown that there are a large number of spin states at low energies in these systems,
thus we can expect a large number of states to contribute in the DMRG-SISO procedure.
Table \ref{tab:tab6} shows how $g$-values change with the number of doublet, quartet, and hextet states included in
the DMRG-SISO calculations for geometry I.
Table \ref{tab:tab7} presents the $g$-values obtained for all three geometries using the 5 lowest doublet and 5 lowest quartet states (10 states in total). 
To include even more states in the DMRG-SISO, we used the ``ground-state specific'' procedure described in section \ref{sec:theory3}
to compute a large number of excited states without explicitly reoptimizing their renormalized bases.
Using this approach we were able to include up to 10 doublet and 10 quartet states. The effect
of including more states on the $g$-tensor for geometry II is presented in Table \ref{tab:tab8}. We see that after 10 doublet and 10 quartet states,
the $g$-tensor appears well converged; the remaining uncertainty is on the $O(0.01)$ level.

\begin{table}[h]
\begin{threeparttable}
\caption{Dependence of $g$-values of the reduced [2Fe-2S] complex on the number of electronic states included in the doublet and quartet manifolds (for geometry II). 
The DMRG energies were converged to better than 10$^{-3} E_h$ accuracy using $M$=3000  for our standard state-specific procedure and using $M$=3200 for the cheap ground-state specific procedure.}
\label{tab:tab8}
\begin{tabular}{m{2cm}m{2cm}m{2cm}m{2cm}m{2cm}m{2cm}m{2cm}}
\hline \hline
  $g$-values   &\multicolumn{2}{c}{3 + 3}&\multicolumn{2}{c}{5 +5}&\multicolumn{2}{c}{10+10}\\
                     & state-specific & g.s. specific & state-specific & g.s. specific & state-specific & g.s. specific\\
\hline
$g_x$                & 1.909&1.907&1.834&1.831&N/A&1.831\\                     
$g_y $               & 1.959&1.953&1.945&1.935&N/A&1.935\\
$g_z$                & 2.004&2.004&1.957&1.962&N/A&1.961\\
\hline \hline
\end{tabular}
\end{threeparttable}
\end{table}

\begin{table}[h]
\begin{threeparttable}
\caption{$g$-values of the reduced \ce{[2Fe-2S]} complex from  DMRG-SISO calculations for all geometries using the 5 lowest doublet and 5 lowest quartet states (10 states in total) and from experiment. The DMRG energies were converged using $M$=3000.}
\label{tab:tab7}
\begin{tabular}{ c|m{2cm}m{2cm}m{2cm}|cccc }
\hline \hline
 $g$-values   & \multicolumn{3}{c|}{Theory} & \multicolumn{4}{c}{Experiment}\\
                    &   \multicolumn{3}{c|}{\ce{[Fe2S2(SCH3)4]^3-}}& &\multicolumn{3}{c}{reduced ferredoxin}\\
                      &  I    &II    &III        & \ce{[Fe2S2(SPh)4]^3-}  &\textit{Anabaena} \cite{Cheng}&Spinach $^a$&\textit{Clostridium}$^b$\\
\hline
$g_x$                & 1.807&1.834 &1.852  &1.87-1.91&1.88&1.86-1.89&1.89-1.93\\                    
$g_y $               & 1.931&1.945 &1.936  &1.91-1.95& 1.96&1.94-1.96&1.95-1.96\\
$g_z$                & 1.969&1.957& 1.964  &2.00-2.01&2.05&2.04-2.05&2.00-2.01\\
\hline \hline
\end{tabular}
\begin{tablenotes}
\item [a] The range is given based on $g$-values for Spinach ferredoxin presented in Ref. \cite{fritz,Fe2S2_g,  gayda3,Malkin,gibson,Mukai}.
\item [b] The range is given based on $g$-values for \textit{Clostridium pasteurianum} ferredoxin in Ref. \cite{Fe2S2_g,Palmer_g}.  
\end{tablenotes}
\end{threeparttable}
\end{table}

Comparing the $g$-tensors from the model DMRG-SISO calculations, and the experimental $g$-tensors in biological complexes in Table~\ref{tab:tab7}, we find that while the middle $g$-value
is in reasonable agreement with experiment, the other two $g$-values are significantly underestimated. As
we have argued, we do not think this is due to insufficient states in the DMRG-SISO procedure. Further, our earlier work has suggested that
the lowest spin state excitation energies are at least qualitatively reasonable in the active space.
We have computed the partial charges on the Fe and S atoms in the DMRG-SISO calculation as well as with the BP86 functional (see Table \ref{tab:chg}). As can be seen the DMRG predicts significantly more ionic Fe--S bonds than at the DFT level.
This suggests that the error in the $g$-values may once again arise from errors in the density and ionicity of the metal-ligand bonds, similar to the case of \ce{CuCl_4^2-} above. 

\begin{table}[h!]
\caption{Selected L{\"{o}}wdin partial charges for the \ce{[Fe2S2(SCH3)4]^3-} complex.}
\label{tab:chg}
\begin{tabular}{ crrrr }
\hline \hline
Atom & \multicolumn{2}{c}{geometry I}& \multicolumn{2}{c}{geometry II}\\
     & BP86&DMRG &BP86&DMRG\\
\hline
Fe1 &0.705&1.097&0.690&0.854\\
Fe2 &0.776&1.047&0.736&0.994\\
S1 &-0.782&-0.910&-0.790&-0.927\\
S2&-0.811&-0.942&-0.801&-0.936\\
\hline \hline
\end{tabular}
\end{table}

\subsection*{D. \ce{Mn4CaO5} model of the S2 state of OEC}

Here we consider a model of the S2 state of the oxygen evolving complex in photosystem II.
We use an oxygen-bridged tetramanganese calcium \ce{Mn4CaO5} complex using the geometry in 
Ref. \cite{oec}, which was optimized for the S2 state using broken-symmetry DFT at the def2-TZVP(-f)/BP86-D3 level of theory,
and with the zeroth-order regular approximation (ZORA) to include scalar relativistic effects. This model has
previously been studied using DMRG in Ref.~\cite{kurashige2013entangled}.


We first carried out an unrestricted BP86/def2-TZVPP-DKH basis set calculation on the high spin state with $S$ = 13/2 and we included scalar-relativistic effects using the X2C method.
As in the previous example, from the alpha and beta UKS orbitals, we constructed  UNOs, 
which were further separated into three subspaces: doubly occupied, singly occupied and virtual molecular orbitals. 
Next, we constructed localized orbitals by projecting atomic orbitals and 
chose the $2p$ orbitals of the five bridging oxygens and $3d$ orbitals of
the four manganese centers
to comprise the active space.
With this (43e, 35o) active space, we calculated 7 doublet and 11 quartet states using DMRG-CI with  $M$=1000.
(Previous studies in Ref.~\cite{kurashige2013entangled} showed that the DMRG energy can be converged
to beyond chemical accuracy at this bond dimension).
We obtained  $g$-values of 2.0014484, 2.0014628, 2.0022972, giving (small) $g$-shifts relative to
the Land{\'e} factor of -870, -856 and -22 ppm.
We are not aware of other theoretical estimates for these $g$-values. However,
in the experimental EPR spectrum of the OEC S2 state with $S$=1/2, one observes
a multiline EPR signal centered at $g$=2.0 \cite{oec,oec2, oec3,oec4}.
There is evidence also that this signal is quite isotropic~\cite{oec,oec2},
and this is consistent with the nearly isotropic $g$-tensor that we compute.

\section{Conclusions}

In this work, we presented a method to calculate molecular $g$-tensors using
state-interaction spin-orbit coupling and density matrix renormalization group wavefunctions.
We have demonstrated this approach on two mononuclear transition
metal complexes and a binuclear and tetranuclear transition metal complexes.
Our results show that it is possible to converge the calculations with respect to the number
of states entering in the state-interaction picture. Remaining discrepancies often appear attributable
to the description of the ionic/covalent character of the metal ligand bond, which requires a careful
balance between static and dynamic correlation. Nonetheless, our work is a step towards truly
multireference calculations of $g$-tensors in complex systems, including in the study of larger active sites in metalloenzymes.

\section*{Acknowledgments}
We acknowledge the US National Science Foundation for funding this research through the award NSF:CHE-1657286.

\bibliography{refs}

\begin{thebibliography}{71}%
\makeatletter
\providecommand \@ifxundefined [1]{%
 \@ifx{#1\undefined}
}%
\providecommand \@ifnum [1]{%
 \ifnum #1\expandafter \@firstoftwo
 \else \expandafter \@secondoftwo
 \fi
}%
\providecommand \@ifx [1]{%
 \ifx #1\expandafter \@firstoftwo
 \else \expandafter \@secondoftwo
 \fi
}%
\providecommand \natexlab [1]{#1}%
\providecommand \enquote  [1]{``#1''}%
\providecommand \bibnamefont  [1]{#1}%
\providecommand \bibfnamefont [1]{#1}%
\providecommand \citenamefont [1]{#1}%
\providecommand \href@noop [0]{\@secondoftwo}%
\providecommand \href [0]{\begingroup \@sanitize@url \@href}%
\providecommand \@href[1]{\@@startlink{#1}\@@href}%
\providecommand \@@href[1]{\endgroup#1\@@endlink}%
\providecommand \@sanitize@url [0]{\catcode `\\12\catcode `\$12\catcode
  `\&12\catcode `\#12\catcode `\^12\catcode `\_12\catcode `\%12\relax}%
\providecommand \@@startlink[1]{}%
\providecommand \@@endlink[0]{}%
\providecommand \url  [0]{\begingroup\@sanitize@url \@url }%
\providecommand \@url [1]{\endgroup\@href {#1}{\urlprefix }}%
\providecommand \urlprefix  [0]{URL }%
\providecommand \Eprint [0]{\href }%
\providecommand \doibase [0]{http://dx.doi.org/}%
\providecommand \selectlanguage [0]{\@gobble}%
\providecommand \bibinfo  [0]{\@secondoftwo}%
\providecommand \bibfield  [0]{\@secondoftwo}%
\providecommand \translation [1]{[#1]}%
\providecommand \BibitemOpen [0]{}%
\providecommand \bibitemStop [0]{}%
\providecommand \bibitemNoStop [0]{.\EOS\space}%
\providecommand \EOS [0]{\spacefactor3000\relax}%
\providecommand \BibitemShut  [1]{\csname bibitem#1\endcsname}%
\let\auto@bib@innerbib\@empty
\bibitem [{\citenamefont {van Lenthe}, \citenamefont {Wormer},\ and\
  \citenamefont {van~der Avoird}(1997)}]{vLenthe}%
  \BibitemOpen
  \bibfield  {author} {\bibinfo {author} {\bibfnamefont {E.}~\bibnamefont {van
  Lenthe}}, \bibinfo {author} {\bibfnamefont {P.~E.~S.}\ \bibnamefont
  {Wormer}}, \ and\ \bibinfo {author} {\bibfnamefont {A.}~\bibnamefont {van~der
  Avoird}},\ }\href {\doibase 10.1063/1.474590} {\bibfield  {journal} {\bibinfo
   {journal} {J. Chem. Phys.}\ }\textbf {\bibinfo {volume} {107}},\ \bibinfo
  {pages} {2488} (\bibinfo {year} {1997})}\BibitemShut {NoStop}%
\bibitem [{\citenamefont {Schreckenbach}\ and\ \citenamefont
  {Ziegler}(1997)}]{ziegler1}%
  \BibitemOpen
  \bibfield  {author} {\bibinfo {author} {\bibfnamefont {G.}~\bibnamefont
  {Schreckenbach}}\ and\ \bibinfo {author} {\bibfnamefont {T.}~\bibnamefont
  {Ziegler}},\ }\href {\doibase 10.1021/jp963060t} {\bibfield  {journal}
  {\bibinfo  {journal} {J. Phys. Chem. A}\ }\textbf {\bibinfo {volume} {101}},\
  \bibinfo {pages} {3388} (\bibinfo {year} {1997})}\BibitemShut {NoStop}%
\bibitem [{\citenamefont {Patchkovskii}\ and\ \citenamefont
  {Ziegler}(2001)}]{ziegler2}%
  \BibitemOpen
  \bibfield  {author} {\bibinfo {author} {\bibfnamefont {S.}~\bibnamefont
  {Patchkovskii}}\ and\ \bibinfo {author} {\bibfnamefont {T.}~\bibnamefont
  {Ziegler}},\ }\href {\doibase 10.1021/jp010457a} {\bibfield  {journal}
  {\bibinfo  {journal} {J. Phys. Chem. A}\ }\textbf {\bibinfo {volume} {105}},\
  \bibinfo {pages} {5490} (\bibinfo {year} {2001})}\BibitemShut {NoStop}%
\bibitem [{\citenamefont {Malkina}\ \emph {et~al.}(2000)\citenamefont
  {Malkina}, \citenamefont {Vaara}, \citenamefont {Schimmelpfenning},
  \citenamefont {Munzarova}, \citenamefont {Malkin},\ and\ \citenamefont
  {Kaupp}}]{Malkina}%
  \BibitemOpen
  \bibfield  {author} {\bibinfo {author} {\bibfnamefont {O.~L.}\ \bibnamefont
  {Malkina}}, \bibinfo {author} {\bibfnamefont {J.}~\bibnamefont {Vaara}},
  \bibinfo {author} {\bibfnamefont {B.}~\bibnamefont {Schimmelpfenning}},
  \bibinfo {author} {\bibfnamefont {M.}~\bibnamefont {Munzarova}}, \bibinfo
  {author} {\bibfnamefont {V.}~\bibnamefont {Malkin}}, \ and\ \bibinfo {author}
  {\bibfnamefont {M.}~\bibnamefont {Kaupp}},\ }\href {\doibase
  10.1021/ja000984s} {\bibfield  {journal} {\bibinfo  {journal} {J. Am. Chem.
  Soc.}\ }\textbf {\bibinfo {volume} {122}},\ \bibinfo {pages} {9206} (\bibinfo
  {year} {2000})}\BibitemShut {NoStop}%
\bibitem [{\citenamefont {Kaupp}\ \emph {et~al.}(2002)\citenamefont {Kaupp},
  \citenamefont {Reviakine}, \citenamefont {Malkina}, \citenamefont
  {Arbuznikov}, \citenamefont {Schimmelpfennig},\ and\ \citenamefont
  {Malkin}}]{Kaupp}%
  \BibitemOpen
  \bibfield  {author} {\bibinfo {author} {\bibfnamefont {M.}~\bibnamefont
  {Kaupp}}, \bibinfo {author} {\bibfnamefont {R.}~\bibnamefont {Reviakine}},
  \bibinfo {author} {\bibfnamefont {O.~L.}\ \bibnamefont {Malkina}}, \bibinfo
  {author} {\bibfnamefont {A.}~\bibnamefont {Arbuznikov}}, \bibinfo {author}
  {\bibfnamefont {B.}~\bibnamefont {Schimmelpfennig}}, \ and\ \bibinfo {author}
  {\bibfnamefont {V.~G.}\ \bibnamefont {Malkin}},\ }\href {\doibase
  10.1002/jcc.10049} {\bibfield  {journal} {\bibinfo  {journal} {J. Comput.
  Chem.}\ }\textbf {\bibinfo {volume} {23}},\ \bibinfo {pages} {794} (\bibinfo
  {year} {2002})}\BibitemShut {NoStop}%
\bibitem [{\citenamefont {Neese}(2001{\natexlab{a}})}]{Neese01_2}%
  \BibitemOpen
  \bibfield  {author} {\bibinfo {author} {\bibfnamefont {F.}~\bibnamefont
  {Neese}},\ }\href {\doibase 10.1063/1.1419058} {\bibfield  {journal}
  {\bibinfo  {journal} {J. Chem. Phys.}\ }\textbf {\bibinfo {volume} {115}},\
  \bibinfo {pages} {11080} (\bibinfo {year} {2001}{\natexlab{a}})}\BibitemShut
  {NoStop}%
\bibitem [{\citenamefont {Neyman}\ \emph {et~al.}(2002)\citenamefont {Neyman},
  \citenamefont {Ganyushin}, \citenamefont {Matveev},\ and\ \citenamefont
  {Nasluzov}}]{g_DKH}%
  \BibitemOpen
  \bibfield  {author} {\bibinfo {author} {\bibfnamefont {K.~M.}\ \bibnamefont
  {Neyman}}, \bibinfo {author} {\bibfnamefont {D.~I.}\ \bibnamefont
  {Ganyushin}}, \bibinfo {author} {\bibfnamefont {A.~V.}\ \bibnamefont
  {Matveev}}, \ and\ \bibinfo {author} {\bibfnamefont {V.~A.}\ \bibnamefont
  {Nasluzov}},\ }\href {\doibase 10.1021/jp0204253} {\bibfield  {journal}
  {\bibinfo  {journal} {J. Phys. Chem. A}\ }\textbf {\bibinfo {volume} {106}},\
  \bibinfo {pages} {5022} (\bibinfo {year} {2002})}\BibitemShut {NoStop}%
\bibitem [{\citenamefont {Malkin}\ \emph {et~al.}(2005)\citenamefont {Malkin},
  \citenamefont {Malkina}, \citenamefont {Malkin},\ and\ \citenamefont
  {Kaupp}}]{g_DKH2}%
  \BibitemOpen
  \bibfield  {author} {\bibinfo {author} {\bibfnamefont {I.}~\bibnamefont
  {Malkin}}, \bibinfo {author} {\bibfnamefont {O.}~\bibnamefont {Malkina}},
  \bibinfo {author} {\bibfnamefont {V.}~\bibnamefont {Malkin}}, \ and\ \bibinfo
  {author} {\bibfnamefont {M.}~\bibnamefont {Kaupp}},\ }\href {\doibase
  10.1063/1.2135290} {\bibfield  {journal} {\bibinfo  {journal} {J. Chem.
  Phys.}\ }\textbf {\bibinfo {volume} {123}},\ \bibinfo {pages} {244103}
  (\bibinfo {year} {2005})}\BibitemShut {NoStop}%
\bibitem [{\citenamefont {Komorovsk{\'y}}\ \emph {et~al.}(2006)\citenamefont
  {Komorovsk{\'y}}, \citenamefont {Repisk{\'y}}, \citenamefont {Malkina},
  \citenamefont {Malkin}, \citenamefont {Malkin},\ and\ \citenamefont
  {Kaupp}}]{Komorovsky}%
  \BibitemOpen
  \bibfield  {author} {\bibinfo {author} {\bibfnamefont {S.}~\bibnamefont
  {Komorovsk{\'y}}}, \bibinfo {author} {\bibfnamefont {M.}~\bibnamefont
  {Repisk{\'y}}}, \bibinfo {author} {\bibfnamefont {O.~L.}\ \bibnamefont
  {Malkina}}, \bibinfo {author} {\bibfnamefont {V.~G.}\ \bibnamefont {Malkin}},
  \bibinfo {author} {\bibfnamefont {I.}~\bibnamefont {Malkin}}, \ and\ \bibinfo
  {author} {\bibfnamefont {M.}~\bibnamefont {Kaupp}},\ }\href {\doibase
  10.1063/1.2173995} {\bibfield  {journal} {\bibinfo  {journal} {J. Chem.
  Phys.}\ }\textbf {\bibinfo {volume} {124}},\ \bibinfo {pages} {084108}
  (\bibinfo {year} {2006})}\BibitemShut {NoStop}%
\bibitem [{\citenamefont {Repisk{\'y}}\ \emph {et~al.}(2010)\citenamefont
  {Repisk{\'y}}, \citenamefont {Komorovsk{\'y}}, \citenamefont {Malkin},
  \citenamefont {Malkina},\ and\ \citenamefont {Malkin}}]{Repisky}%
  \BibitemOpen
  \bibfield  {author} {\bibinfo {author} {\bibfnamefont {M.}~\bibnamefont
  {Repisk{\'y}}}, \bibinfo {author} {\bibfnamefont {S.}~\bibnamefont
  {Komorovsk{\'y}}}, \bibinfo {author} {\bibfnamefont {E.}~\bibnamefont
  {Malkin}}, \bibinfo {author} {\bibfnamefont {O.~L.}\ \bibnamefont {Malkina}},
  \ and\ \bibinfo {author} {\bibfnamefont {V.~G.}\ \bibnamefont {Malkin}},\
  }\href {\doibase 10.1016/j.cplett.2010.01.077} {\bibfield  {journal}
  {\bibinfo  {journal} {Chem. Phys. Lett.}\ }\textbf {\bibinfo {volume}
  {488}},\ \bibinfo {pages} {94} (\bibinfo {year} {2010})}\BibitemShut
  {NoStop}%
\bibitem [{\citenamefont {Rinkevicius}\ \emph {et~al.}(2003)\citenamefont
  {Rinkevicius}, \citenamefont {Telyatnyk}, \citenamefont {Sa{\l}ek},
  \citenamefont {Vahtras},\ and\ \citenamefont {{\AA}gren}}]{Rinkevicius}%
  \BibitemOpen
  \bibfield  {author} {\bibinfo {author} {\bibfnamefont {Z.}~\bibnamefont
  {Rinkevicius}}, \bibinfo {author} {\bibfnamefont {L.}~\bibnamefont
  {Telyatnyk}}, \bibinfo {author} {\bibfnamefont {P.}~\bibnamefont {Sa{\l}ek}},
  \bibinfo {author} {\bibfnamefont {O.}~\bibnamefont {Vahtras}}, \ and\
  \bibinfo {author} {\bibfnamefont {H.}~\bibnamefont {{\AA}gren}},\ }\href
  {\doibase 10.1063/1.1620497} {\bibfield  {journal} {\bibinfo  {journal} {J.
  Chem. Phys.}\ }\textbf {\bibinfo {volume} {119}},\ \bibinfo {pages} {10489}
  (\bibinfo {year} {2003})}\BibitemShut {NoStop}%
\bibitem [{\citenamefont {Lushington}, \citenamefont {B{\"u}ndgen},\ and\
  \citenamefont {Grein}(1995)}]{Lushington1}%
  \BibitemOpen
  \bibfield  {author} {\bibinfo {author} {\bibfnamefont {G.}~\bibnamefont
  {Lushington}}, \bibinfo {author} {\bibfnamefont {P.}~\bibnamefont
  {B{\"u}ndgen}}, \ and\ \bibinfo {author} {\bibfnamefont {F.}~\bibnamefont
  {Grein}},\ }\href {\doibase 10.1002/qua.560550503} {\bibfield  {journal}
  {\bibinfo  {journal} {Int. J. Quantum Chem.}\ }\textbf {\bibinfo {volume}
  {55}},\ \bibinfo {pages} {377} (\bibinfo {year} {1995})}\BibitemShut
  {NoStop}%
\bibitem [{\citenamefont {B{\"u}ndgen}, \citenamefont {Lushington},\ and\
  \citenamefont {Grein}(1995)}]{Bundgen}%
  \BibitemOpen
  \bibfield  {author} {\bibinfo {author} {\bibfnamefont {P.}~\bibnamefont
  {B{\"u}ndgen}}, \bibinfo {author} {\bibfnamefont {G.}~\bibnamefont
  {Lushington}}, \ and\ \bibinfo {author} {\bibfnamefont {F.}~\bibnamefont
  {Grein}},\ }\href {\doibase 10.1002/qua.560560831} {\bibfield  {journal}
  {\bibinfo  {journal} {Int. J. Quantum Chem.}\ }\textbf {\bibinfo {volume}
  {56}},\ \bibinfo {pages} {283} (\bibinfo {year} {1995})}\BibitemShut
  {NoStop}%
\bibitem [{\citenamefont {Lushington}\ and\ \citenamefont
  {Grein}(1996)}]{Lushington2}%
  \BibitemOpen
  \bibfield  {author} {\bibinfo {author} {\bibfnamefont {G.~H.}\ \bibnamefont
  {Lushington}}\ and\ \bibinfo {author} {\bibfnamefont {F.}~\bibnamefont
  {Grein}},\ }\href {\doibase 10.1007/BF01127505} {\bibfield  {journal}
  {\bibinfo  {journal} {Theor. Chim. Act.}\ }\textbf {\bibinfo {volume} {93}},\
  \bibinfo {pages} {259} (\bibinfo {year} {1996})}\BibitemShut {NoStop}%
\bibitem [{\citenamefont {Bruna}, \citenamefont {Lushington},\ and\
  \citenamefont {Grein}(1997)}]{Bruna}%
  \BibitemOpen
  \bibfield  {author} {\bibinfo {author} {\bibfnamefont {P.~J.}\ \bibnamefont
  {Bruna}}, \bibinfo {author} {\bibfnamefont {G.~H.}\ \bibnamefont
  {Lushington}}, \ and\ \bibinfo {author} {\bibfnamefont {F.}~\bibnamefont
  {Grein}},\ }\href {\doibase 10.1016/S0301-0104(97)00250-4} {\bibfield
  {journal} {\bibinfo  {journal} {Chem. Phys.}\ }\textbf {\bibinfo {volume}
  {225}},\ \bibinfo {pages} {1} (\bibinfo {year} {1997})}\BibitemShut {NoStop}%
\bibitem [{\citenamefont {Neese}(2003)}]{Neese03}%
  \BibitemOpen
  \bibfield  {author} {\bibinfo {author} {\bibfnamefont {F.}~\bibnamefont
  {Neese}},\ }\href {\doibase 10.1016/j.cplett.2003.09.047} {\bibfield
  {journal} {\bibinfo  {journal} {Chem. Phys. Lett.}\ }\textbf {\bibinfo
  {volume} {380}},\ \bibinfo {pages} {721} (\bibinfo {year}
  {2003})}\BibitemShut {NoStop}%
\bibitem [{\citenamefont {Brownridge}\ \emph {et~al.}(2003)\citenamefont
  {Brownridge}, \citenamefont {Grein}, \citenamefont {Tatchen}, \citenamefont
  {Kleinschmidt},\ and\ \citenamefont {Marian}}]{Brownridge}%
  \BibitemOpen
  \bibfield  {author} {\bibinfo {author} {\bibfnamefont {S.}~\bibnamefont
  {Brownridge}}, \bibinfo {author} {\bibfnamefont {F.}~\bibnamefont {Grein}},
  \bibinfo {author} {\bibfnamefont {J.}~\bibnamefont {Tatchen}}, \bibinfo
  {author} {\bibfnamefont {M.}~\bibnamefont {Kleinschmidt}}, \ and\ \bibinfo
  {author} {\bibfnamefont {C.}~\bibnamefont {Marian}},\ }\href {\doibase
  10.1063/1.1569243} {\bibfield  {journal} {\bibinfo  {journal} {J. Chem.
  Phys.}\ }\textbf {\bibinfo {volume} {118}},\ \bibinfo {pages} {9552}
  (\bibinfo {year} {2003})}\BibitemShut {NoStop}%
\bibitem [{\citenamefont {Neese}(2007)}]{Neese07}%
  \BibitemOpen
  \bibfield  {author} {\bibinfo {author} {\bibfnamefont {F.}~\bibnamefont
  {Neese}},\ }\href {\doibase 10.1080/00268970701549389} {\bibfield  {journal}
  {\bibinfo  {journal} {Mol. Phys.}\ }\textbf {\bibinfo {volume} {105}},\
  \bibinfo {pages} {2507} (\bibinfo {year} {2007})}\BibitemShut {NoStop}%
\bibitem [{\citenamefont {Neese}(2001{\natexlab{b}})}]{Neese01}%
  \BibitemOpen
  \bibfield  {author} {\bibinfo {author} {\bibfnamefont {F.}~\bibnamefont
  {Neese}},\ }\href {\doibase 10.1002/qua.1202} {\bibfield  {journal} {\bibinfo
   {journal} {Int. J. Quant. Chem.}\ }\textbf {\bibinfo {volume} {83}},\
  \bibinfo {pages} {104} (\bibinfo {year} {2001}{\natexlab{b}})}\BibitemShut
  {NoStop}%
\bibitem [{\citenamefont {Tatchen}, \citenamefont {Kleinschmidt},\ and\
  \citenamefont {Marian}(2009)}]{gSOCI}%
  \BibitemOpen
  \bibfield  {author} {\bibinfo {author} {\bibfnamefont {J.}~\bibnamefont
  {Tatchen}}, \bibinfo {author} {\bibfnamefont {M.}~\bibnamefont
  {Kleinschmidt}}, \ and\ \bibinfo {author} {\bibfnamefont {C.}~\bibnamefont
  {Marian}},\ }\href {\doibase 10.1063/1.3115965} {\bibfield  {journal}
  {\bibinfo  {journal} {J. Chem. Phys.}\ }\textbf {\bibinfo {volume} {130}},\
  \bibinfo {pages} {154106} (\bibinfo {year} {2009})}\BibitemShut {NoStop}%
\bibitem [{\citenamefont {Vahtras}, \citenamefont {Minaev},\ and\ \citenamefont
  {{\AA}gren}(1997)}]{Vahtras}%
  \BibitemOpen
  \bibfield  {author} {\bibinfo {author} {\bibfnamefont {O.}~\bibnamefont
  {Vahtras}}, \bibinfo {author} {\bibfnamefont {B.}~\bibnamefont {Minaev}}, \
  and\ \bibinfo {author} {\bibfnamefont {H.}~\bibnamefont {{\AA}gren}},\ }\href
  {\doibase 10.1016/S0009-2614(97)01169-X} {\bibfield  {journal} {\bibinfo
  {journal} {Chem. Phys. Lett.}\ }\textbf {\bibinfo {volume} {281}},\ \bibinfo
  {pages} {186} (\bibinfo {year} {1997})}\BibitemShut {NoStop}%
\bibitem [{\citenamefont {Bolvin}(2006)}]{Bolvin}%
  \BibitemOpen
  \bibfield  {author} {\bibinfo {author} {\bibfnamefont {H.}~\bibnamefont
  {Bolvin}},\ }\href {\doibase 10.1002/cphc.200600051} {\bibfield  {journal}
  {\bibinfo  {journal} {Chem. Phys. Chem.}\ }\textbf {\bibinfo {volume} {7}},\
  \bibinfo {pages} {1575} (\bibinfo {year} {2006})}\BibitemShut {NoStop}%
\bibitem [{\citenamefont {Ganyushin}\ and\ \citenamefont
  {Neese}(2013)}]{Ganyushin}%
  \BibitemOpen
  \bibfield  {author} {\bibinfo {author} {\bibfnamefont {D.}~\bibnamefont
  {Ganyushin}}\ and\ \bibinfo {author} {\bibfnamefont {F.}~\bibnamefont
  {Neese}},\ }\href {\doibase 10.1063/1.4793736} {\bibfield  {journal}
  {\bibinfo  {journal} {J. Chem. Phys.}\ }\textbf {\bibinfo {volume} {138}},\
  \bibinfo {pages} {104113} (\bibinfo {year} {2013})}\BibitemShut {NoStop}%
\bibitem [{\citenamefont {Lan}, \citenamefont {Chalupsk{\'y}},\ and\
  \citenamefont {Yanai}(2015)}]{Lan}%
  \BibitemOpen
  \bibfield  {author} {\bibinfo {author} {\bibfnamefont {T.~N.}\ \bibnamefont
  {Lan}}, \bibinfo {author} {\bibfnamefont {J.}~\bibnamefont {Chalupsk{\'y}}},
  \ and\ \bibinfo {author} {\bibfnamefont {T.}~\bibnamefont {Yanai}},\ }\href
  {\doibase 10.1080/00268976.2015.1012128} {\bibfield  {journal} {\bibinfo
  {journal} {Mol. Phys.}\ }\textbf {\bibinfo {volume} {113}},\ \bibinfo {pages}
  {1750} (\bibinfo {year} {2015})}\BibitemShut {NoStop}%
\bibitem [{\citenamefont {Vancoillie}, \citenamefont {Malmqvist},\ and\
  \citenamefont {Pierloot}(2007)}]{Vancoillie}%
  \BibitemOpen
  \bibfield  {author} {\bibinfo {author} {\bibfnamefont {S.}~\bibnamefont
  {Vancoillie}}, \bibinfo {author} {\bibfnamefont {P.-{\AA}.}\ \bibnamefont
  {Malmqvist}}, \ and\ \bibinfo {author} {\bibfnamefont {K.}~\bibnamefont
  {Pierloot}},\ }\href {\doibase 10.1002/cphc.200700128} {\bibfield  {journal}
  {\bibinfo  {journal} {ChemPhysChem}\ }\textbf {\bibinfo {volume} {8}},\
  \bibinfo {pages} {1803} (\bibinfo {year} {2007})}\BibitemShut {NoStop}%
\bibitem [{\citenamefont {Gauss}, \citenamefont {K{\'a}llay},\ and\
  \citenamefont {Neese}(2009)}]{Gauss}%
  \BibitemOpen
  \bibfield  {author} {\bibinfo {author} {\bibfnamefont {J.}~\bibnamefont
  {Gauss}}, \bibinfo {author} {\bibfnamefont {M.}~\bibnamefont {K{\'a}llay}}, \
  and\ \bibinfo {author} {\bibfnamefont {F.}~\bibnamefont {Neese}},\ }\href
  {\doibase 10.1021/jp9028535} {\bibfield  {journal} {\bibinfo  {journal} {J.
  Phys. Chem. A}\ }\textbf {\bibinfo {volume} {113}},\ \bibinfo {pages} {11541}
  (\bibinfo {year} {2009})}\BibitemShut {NoStop}%
\bibitem [{\citenamefont {Roemelt}(2015)}]{roemelt}%
  \BibitemOpen
  \bibfield  {author} {\bibinfo {author} {\bibfnamefont {M.}~\bibnamefont
  {Roemelt}},\ }\href {\doibase 10.1063/1.4927432} {\bibfield  {journal}
  {\bibinfo  {journal} {J. Chem. Phys.}\ }\textbf {\bibinfo {volume} {143}},\
  \bibinfo {pages} {044112} (\bibinfo {year} {2015})}\BibitemShut {NoStop}%
\bibitem [{\citenamefont {Sayfutyarova}\ and\ \citenamefont
  {Chan}(2016)}]{elviras}%
  \BibitemOpen
  \bibfield  {author} {\bibinfo {author} {\bibfnamefont {E.~R.}\ \bibnamefont
  {Sayfutyarova}}\ and\ \bibinfo {author} {\bibfnamefont {G.~K.-L.}\
  \bibnamefont {Chan}},\ }\href {\doibase 10.1063/1.4953445} {\bibfield
  {journal} {\bibinfo  {journal} {J. Chem. Phys.}\ }\textbf {\bibinfo {volume}
  {144}},\ \bibinfo {pages} {234301} (\bibinfo {year} {2016})}\BibitemShut
  {NoStop}%
\bibitem [{\citenamefont {Knecht}\ \emph {et~al.}(2016)\citenamefont {Knecht},
  \citenamefont {Keller}, \citenamefont {Autschbach},\ and\ \citenamefont
  {Reiher}}]{reiher}%
  \BibitemOpen
  \bibfield  {author} {\bibinfo {author} {\bibfnamefont {S.}~\bibnamefont
  {Knecht}}, \bibinfo {author} {\bibfnamefont {S.}~\bibnamefont {Keller}},
  \bibinfo {author} {\bibfnamefont {J.}~\bibnamefont {Autschbach}}, \ and\
  \bibinfo {author} {\bibfnamefont {M.}~\bibnamefont {Reiher}},\ }\href
  {\doibase 10.1021/acs.jctc.6b00889} {\bibfield  {journal} {\bibinfo
  {journal} {J. Chem. Theory Comput.}\ }\textbf {\bibinfo {volume} {12}},\
  \bibinfo {pages} {5881} (\bibinfo {year} {2016})}\BibitemShut {NoStop}%
\bibitem [{\citenamefont {Chibotaru}\ and\ \citenamefont
  {Ungur}(2012)}]{Chibotaru}%
  \BibitemOpen
  \bibfield  {author} {\bibinfo {author} {\bibfnamefont {L.~F.}\ \bibnamefont
  {Chibotaru}}\ and\ \bibinfo {author} {\bibfnamefont {L.}~\bibnamefont
  {Ungur}},\ }\href {\doibase 10.1063/1.4739763} {\bibfield  {journal}
  {\bibinfo  {journal} {J. Chem. Phys.}\ }\textbf {\bibinfo {volume} {137}},\
  \bibinfo {pages} {064112} (\bibinfo {year} {2012})}\BibitemShut {NoStop}%
\bibitem [{\citenamefont {Gerloch}\ and\ \citenamefont
  {McMeeking}(1975)}]{gerloch}%
  \BibitemOpen
  \bibfield  {author} {\bibinfo {author} {\bibfnamefont {M.}~\bibnamefont
  {Gerloch}}\ and\ \bibinfo {author} {\bibfnamefont {R.~F.}\ \bibnamefont
  {McMeeking}},\ }\href {\doibase 10.1039/DT9750002443} {\bibfield  {journal}
  {\bibinfo  {journal} {J. Chem. Soc., Dalton Trans.}\ ,\ \bibinfo {pages}
  {2443}} (\bibinfo {year} {1975})}\BibitemShut {NoStop}%
\bibitem [{\citenamefont {Chibotaru}\ \emph {et~al.}(2005)\citenamefont
  {Chibotaru}, \citenamefont {Hendrickx}, \citenamefont {Clima}, \citenamefont
  {Larionova},\ and\ \citenamefont {Ceulemans}}]{Chibotaru2}%
  \BibitemOpen
  \bibfield  {author} {\bibinfo {author} {\bibfnamefont {L.~F.}\ \bibnamefont
  {Chibotaru}}, \bibinfo {author} {\bibfnamefont {M.~F.~A.}\ \bibnamefont
  {Hendrickx}}, \bibinfo {author} {\bibfnamefont {S.}~\bibnamefont {Clima}},
  \bibinfo {author} {\bibfnamefont {J.}~\bibnamefont {Larionova}}, \ and\
  \bibinfo {author} {\bibfnamefont {A.}~\bibnamefont {Ceulemans}},\ }\href
  {\doibase 10.1021/jp051858j} {\bibfield  {journal} {\bibinfo  {journal} {J.
  Phys. Chem. A}\ }\textbf {\bibinfo {volume} {109}},\ \bibinfo {pages} {7251}
  (\bibinfo {year} {2005})}\BibitemShut {NoStop}%
\bibitem [{\citenamefont {Marian}(2001)}]{Marian:SOC}%
  \BibitemOpen
  \bibfield  {author} {\bibinfo {author} {\bibfnamefont {C.~M.}\ \bibnamefont
  {Marian}},\ }in\ \href {\doibase 10.1002/0471224413.ch3} {\emph {\bibinfo
  {booktitle} {Reviews in Computational Chemistry}}},\ Vol.~\bibinfo {volume}
  {17},\ \bibinfo {editor} {edited by\ \bibinfo {editor} {\bibfnamefont
  {K.~B.}\ \bibnamefont {Lipkowitz}}\ and\ \bibinfo {editor} {\bibfnamefont
  {D.~B.}\ \bibnamefont {Boyd}}}\ (\bibinfo  {publisher} {Wiley-VCH},\ \bibinfo
  {address} {New York},\ \bibinfo {year} {2001})\ pp.\ \bibinfo {pages}
  {99--204}\BibitemShut {NoStop}%
\bibitem [{\citenamefont {He{\ss}}\ \emph {et~al.}(1996)\citenamefont
  {He{\ss}}, \citenamefont {C.M.Marian}, \citenamefont {Wahlgren},\ and\
  \citenamefont {Gropen}}]{Hess:AMFI}%
  \BibitemOpen
  \bibfield  {author} {\bibinfo {author} {\bibfnamefont {B.~A.}\ \bibnamefont
  {He{\ss}}}, \bibinfo {author} {\bibnamefont {C.M.Marian}}, \bibinfo {author}
  {\bibfnamefont {U.}~\bibnamefont {Wahlgren}}, \ and\ \bibinfo {author}
  {\bibfnamefont {O.}~\bibnamefont {Gropen}},\ }\href {\doibase
  10.1016/0009-2614(96)00119-4} {\bibfield  {journal} {\bibinfo  {journal}
  {Chem. Phys. Lett.}\ }\textbf {\bibinfo {volume} {251}},\ \bibinfo {pages}
  {365} (\bibinfo {year} {1996})}\BibitemShut {NoStop}%
\bibitem [{\citenamefont {Tatchen}\ and\ \citenamefont
  {Marian}(1999)}]{Tatchen}%
  \BibitemOpen
  \bibfield  {author} {\bibinfo {author} {\bibfnamefont {J.}~\bibnamefont
  {Tatchen}}\ and\ \bibinfo {author} {\bibfnamefont {C.~M.}\ \bibnamefont
  {Marian}},\ }\href {\doibase 10.1016/S0009-2614(99)00997-5} {\bibfield
  {journal} {\bibinfo  {journal} {Chem. Phys. Lett.}\ }\textbf {\bibinfo
  {volume} {313}},\ \bibinfo {pages} {351} (\bibinfo {year}
  {1999})}\BibitemShut {NoStop}%
\bibitem [{\citenamefont {Neese}(2005)}]{Neese:SOMF}%
  \BibitemOpen
  \bibfield  {author} {\bibinfo {author} {\bibfnamefont {F.}~\bibnamefont
  {Neese}},\ }\href {\doibase 10.1063/1.1829047} {\bibfield  {journal}
  {\bibinfo  {journal} {J. Chem. Phys.}\ }\textbf {\bibinfo {volume} {122}},\
  \bibinfo {pages} {034107} (\bibinfo {year} {2005})}\BibitemShut {NoStop}%
\bibitem [{\citenamefont {Malmqvist}, \citenamefont {Roos},\ and\ \citenamefont
  {Schimmelpfennig}(2002)}]{Malmqvist:RASSI_SO}%
  \BibitemOpen
  \bibfield  {author} {\bibinfo {author} {\bibfnamefont {P.-{\AA}.}\
  \bibnamefont {Malmqvist}}, \bibinfo {author} {\bibfnamefont {B.}~\bibnamefont
  {Roos}}, \ and\ \bibinfo {author} {\bibfnamefont {B.}~\bibnamefont
  {Schimmelpfennig}},\ }\href {\doibase 10.1016/S0009-2614(02)00498-0}
  {\bibfield  {journal} {\bibinfo  {journal} {Chem. Phys. Lett.}\ }\textbf
  {\bibinfo {volume} {357}},\ \bibinfo {pages} {230} (\bibinfo {year}
  {2002})}\BibitemShut {NoStop}%
\bibitem [{\citenamefont {Helgaker}, \citenamefont {J{\o}rgensen},\ and\
  \citenamefont {Olsen}(2000)}]{helgaker:purplebook}%
  \BibitemOpen
  \bibfield  {author} {\bibinfo {author} {\bibfnamefont {T.}~\bibnamefont
  {Helgaker}}, \bibinfo {author} {\bibfnamefont {P.}~\bibnamefont
  {J{\o}rgensen}}, \ and\ \bibinfo {author} {\bibfnamefont {J.}~\bibnamefont
  {Olsen}},\ }\href@noop {} {\emph {\bibinfo {title} {Molecular Electronic
  Structure Theory}}}\ (\bibinfo  {publisher} {Wiley},\ \bibinfo {address}
  {Chichester},\ \bibinfo {year} {2000})\BibitemShut {NoStop}%
\bibitem [{\citenamefont {Helgaker}\ \emph {et~al.}(2012)\citenamefont
  {Helgaker}, \citenamefont {Coriani}, \citenamefont {J{\o}rgensen},
  \citenamefont {Kristensen}, \citenamefont {Olsen},\ and\ \citenamefont
  {Ruud}}]{MolProp}%
  \BibitemOpen
  \bibfield  {author} {\bibinfo {author} {\bibfnamefont {T.}~\bibnamefont
  {Helgaker}}, \bibinfo {author} {\bibfnamefont {S.}~\bibnamefont {Coriani}},
  \bibinfo {author} {\bibfnamefont {P.}~\bibnamefont {J{\o}rgensen}}, \bibinfo
  {author} {\bibfnamefont {K.}~\bibnamefont {Kristensen}}, \bibinfo {author}
  {\bibfnamefont {J.}~\bibnamefont {Olsen}}, \ and\ \bibinfo {author}
  {\bibfnamefont {K.}~\bibnamefont {Ruud}},\ }\href {\doibase
  10.1021/cr2002239} {\bibfield  {journal} {\bibinfo  {journal} {Chem. Rev.}\
  }\textbf {\bibinfo {volume} {112}},\ \bibinfo {pages} {543} (\bibinfo {year}
  {2012})}\BibitemShut {NoStop}%
\bibitem [{\citenamefont {Van Yperen-De~Deyne}\ \emph
  {et~al.}(2012)\citenamefont {Van Yperen-De~Deyne}, \citenamefont {Pauwels},
  \citenamefont {Van~Speybroeck},\ and\ \citenamefont {Waroquier}}]{Waroquier}%
  \BibitemOpen
  \bibfield  {author} {\bibinfo {author} {\bibfnamefont {A.}~\bibnamefont {Van
  Yperen-De~Deyne}}, \bibinfo {author} {\bibfnamefont {E.}~\bibnamefont
  {Pauwels}}, \bibinfo {author} {\bibfnamefont {V.}~\bibnamefont
  {Van~Speybroeck}}, \ and\ \bibinfo {author} {\bibfnamefont {M.}~\bibnamefont
  {Waroquier}},\ }\href {\doibase 10.1039/c2cp41086a} {\bibfield  {journal}
  {\bibinfo  {journal} {Phys. Chem. Chem. Phys}\ }\textbf {\bibinfo {volume}
  {14}},\ \bibinfo {pages} {10690} (\bibinfo {year} {2012})}\BibitemShut
  {NoStop}%
\bibitem [{\citenamefont {Sharma}\ and\ \citenamefont
  {Chan}(2012)}]{Sandeep:dmrg}%
  \BibitemOpen
  \bibfield  {author} {\bibinfo {author} {\bibfnamefont {S.}~\bibnamefont
  {Sharma}}\ and\ \bibinfo {author} {\bibfnamefont {G.~K.-L.}\ \bibnamefont
  {Chan}},\ }\href {\doibase 10.1063/1.3695642} {\bibfield  {journal} {\bibinfo
   {journal} {J. Chem. Phys.}\ }\textbf {\bibinfo {volume} {136}},\ \bibinfo
  {pages} {124121} (\bibinfo {year} {2012})}\BibitemShut {NoStop}%
\bibitem [{\citenamefont {Sun}\ \emph {et~al.}(2017)\citenamefont {Sun},
  \citenamefont {Berkelbach}, \citenamefont {Blunt}, \citenamefont {Booth},
  \citenamefont {Guo}, \citenamefont {Li}, \citenamefont {Liu}, \citenamefont
  {McClain}, \citenamefont {Sayfutyarova}, \citenamefont {Sharma},
  \citenamefont {Wouters},\ and\ \citenamefont {Chan}}]{pyscf}%
  \BibitemOpen
  \bibfield  {author} {\bibinfo {author} {\bibfnamefont {Q.}~\bibnamefont
  {Sun}}, \bibinfo {author} {\bibfnamefont {T.~C.}\ \bibnamefont {Berkelbach}},
  \bibinfo {author} {\bibfnamefont {N.~S.}\ \bibnamefont {Blunt}}, \bibinfo
  {author} {\bibfnamefont {G.~H.}\ \bibnamefont {Booth}}, \bibinfo {author}
  {\bibfnamefont {S.}~\bibnamefont {Guo}}, \bibinfo {author} {\bibfnamefont
  {Z.}~\bibnamefont {Li}}, \bibinfo {author} {\bibfnamefont {J.}~\bibnamefont
  {Liu}}, \bibinfo {author} {\bibfnamefont {J.}~\bibnamefont {McClain}},
  \bibinfo {author} {\bibfnamefont {E.~R.}\ \bibnamefont {Sayfutyarova}},
  \bibinfo {author} {\bibfnamefont {S.}~\bibnamefont {Sharma}}, \bibinfo
  {author} {\bibfnamefont {S.}~\bibnamefont {Wouters}}, \ and\ \bibinfo
  {author} {\bibfnamefont {G.~K.-L.}\ \bibnamefont {Chan}},\ }\href {\doibase
  10.1002/wcms.1340} {\bibfield  {journal} {\bibinfo  {journal} {WIRE: Comput.
  Mol. Sci.}\ }\textbf {\bibinfo {volume} {8}} (\bibinfo {year} {2017}),\
  10.1002/wcms.1340}\BibitemShut {NoStop}%
\bibitem [{\citenamefont {Wolf}, \citenamefont {Reiher},\ and\ \citenamefont
  {Hess}(2002)}]{DKH1}%
  \BibitemOpen
  \bibfield  {author} {\bibinfo {author} {\bibfnamefont {A.}~\bibnamefont
  {Wolf}}, \bibinfo {author} {\bibfnamefont {M.}~\bibnamefont {Reiher}}, \ and\
  \bibinfo {author} {\bibfnamefont {B.~A.}\ \bibnamefont {Hess}},\ }\href
  {\doibase 10.1063/1.1515314} {\bibfield  {journal} {\bibinfo  {journal} {J.
  Chem. Phys.}\ }\textbf {\bibinfo {volume} {117}},\ \bibinfo {pages} {9215}
  (\bibinfo {year} {2002})}\BibitemShut {NoStop}%
\bibitem [{\citenamefont {Reiher}\ and\ \citenamefont
  {Wolf}(2004{\natexlab{a}})}]{DKH2}%
  \BibitemOpen
  \bibfield  {author} {\bibinfo {author} {\bibfnamefont {M.}~\bibnamefont
  {Reiher}}\ and\ \bibinfo {author} {\bibfnamefont {A.}~\bibnamefont {Wolf}},\
  }\href {\doibase 10.1063/1.1768160} {\bibfield  {journal} {\bibinfo
  {journal} {J. Chem. Phys.}\ }\textbf {\bibinfo {volume} {121}},\ \bibinfo
  {pages} {2037} (\bibinfo {year} {2004}{\natexlab{a}})}\BibitemShut {NoStop}%
\bibitem [{\citenamefont {Reiher}\ and\ \citenamefont
  {Wolf}(2004{\natexlab{b}})}]{DKH3}%
  \BibitemOpen
  \bibfield  {author} {\bibinfo {author} {\bibfnamefont {M.}~\bibnamefont
  {Reiher}}\ and\ \bibinfo {author} {\bibfnamefont {A.}~\bibnamefont {Wolf}},\
  }\href {\doibase 10.1063/1.1818681} {\bibfield  {journal} {\bibinfo
  {journal} {J. Chem. Phys.}\ }\textbf {\bibinfo {volume} {121}},\ \bibinfo
  {pages} {10945} (\bibinfo {year} {2004}{\natexlab{b}})}\BibitemShut {NoStop}%
\bibitem [{\citenamefont {Solomonik}\ and\ \citenamefont
  {Mukhanov}(2014)}]{solomonik}%
  \BibitemOpen
  \bibfield  {author} {\bibinfo {author} {\bibfnamefont {V.~G.}\ \bibnamefont
  {Solomonik}}\ and\ \bibinfo {author} {\bibfnamefont {A.~A.}\ \bibnamefont
  {Mukhanov}},\ }\href {\doibase 10.1134/S0036024414010233} {\bibfield
  {journal} {\bibinfo  {journal} {Russ. J. Phys. Chem. A}\ }\textbf {\bibinfo
  {volume} {88}},\ \bibinfo {pages} {85} (\bibinfo {year} {2014})}\BibitemShut
  {NoStop}%
\bibitem [{\citenamefont {DeVore}\ and\ \citenamefont
  {W.~Weltner}(1977)}]{deVore}%
  \BibitemOpen
  \bibfield  {author} {\bibinfo {author} {\bibfnamefont {T.~C.}\ \bibnamefont
  {DeVore}}\ and\ \bibinfo {author} {\bibfnamefont {J.}~\bibnamefont
  {W.~Weltner}},\ }\href {\doibase 10.1021/ja00456a028} {\bibfield  {journal}
  {\bibinfo  {journal} {J. Am. Chem. Soc.}\ }\textbf {\bibinfo {volume} {99}},\
  \bibinfo {pages} {4700} (\bibinfo {year} {1977})}\BibitemShut {NoStop}%
\bibitem [{\citenamefont {Werner}\ \emph {et~al.}(2012)\citenamefont {Werner},
  \citenamefont {Knowles}, \citenamefont {Knizia}, \citenamefont {Manby},\ and\
  \citenamefont {Sch{\"u}tz}}]{molpro}%
  \BibitemOpen
  \bibfield  {author} {\bibinfo {author} {\bibfnamefont {H.-J.}\ \bibnamefont
  {Werner}}, \bibinfo {author} {\bibfnamefont {P.~J.}\ \bibnamefont {Knowles}},
  \bibinfo {author} {\bibfnamefont {G.}~\bibnamefont {Knizia}}, \bibinfo
  {author} {\bibfnamefont {F.~R.}\ \bibnamefont {Manby}}, \ and\ \bibinfo
  {author} {\bibfnamefont {M.}~\bibnamefont {Sch{\"u}tz}},\ }\href {\doibase
  10.1002/wcms.82} {\bibfield  {journal} {\bibinfo  {journal} {WIREs: Comput.
  Mol. Sci.}\ }\textbf {\bibinfo {volume} {2}},\ \bibinfo {pages} {242}
  (\bibinfo {year} {2012})}\BibitemShut {NoStop}%
\bibitem [{\citenamefont {Willett}, \citenamefont {Liles},\ and\ \citenamefont
  {Michelson}(1967)}]{Willet}%
  \BibitemOpen
  \bibfield  {author} {\bibinfo {author} {\bibfnamefont {R.~D.}\ \bibnamefont
  {Willett}}, \bibinfo {author} {\bibfnamefont {O.~L.}\ \bibnamefont {Liles}},
  \ and\ \bibinfo {author} {\bibfnamefont {C.}~\bibnamefont {Michelson}},\
  }\href {\doibase 10.1021/ic50056a028} {\bibfield  {journal} {\bibinfo
  {journal} {Inorg. Chem.}\ }\textbf {\bibinfo {volume} {6}},\ \bibinfo {pages}
  {1885} (\bibinfo {year} {1967})}\BibitemShut {NoStop}%
\bibitem [{\citenamefont {Solomon}\ \emph {et~al.}(2004)\citenamefont
  {Solomon}, \citenamefont {Szilagyi}, \citenamefont {George},\ and\
  \citenamefont {Basumallick}}]{Solomon1}%
  \BibitemOpen
  \bibfield  {author} {\bibinfo {author} {\bibfnamefont {E.~I.}\ \bibnamefont
  {Solomon}}, \bibinfo {author} {\bibfnamefont {R.~K.}\ \bibnamefont
  {Szilagyi}}, \bibinfo {author} {\bibfnamefont {S.~D.}\ \bibnamefont
  {George}}, \ and\ \bibinfo {author} {\bibfnamefont {L.}~\bibnamefont
  {Basumallick}},\ }\href {\doibase 10.1021/cr0206317} {\bibfield  {journal}
  {\bibinfo  {journal} {Chem. Rev.}\ }\textbf {\bibinfo {volume} {104}},\
  \bibinfo {pages} {419} (\bibinfo {year} {2004})}\BibitemShut {NoStop}%
\bibitem [{\citenamefont {Neese}\ and\ \citenamefont
  {Solomon}(2002)}]{Solomon2}%
  \BibitemOpen
  \bibfield  {author} {\bibinfo {author} {\bibfnamefont {F.}~\bibnamefont
  {Neese}}\ and\ \bibinfo {author} {\bibfnamefont {E.~I.}\ \bibnamefont
  {Solomon}},\ }\enquote {\bibinfo {title} {Interpretation and calculation of
  spin-hamiltonian parameters in transition metal complexes},}\ in\ \href
  {\doibase 10.1002/3527600698.ch9} {\emph {\bibinfo {booktitle} {Magnetism:
  Molecules to Materials IV: Nanosized Magnetic Materials}}},\ \bibinfo
  {editor} {edited by\ \bibinfo {editor} {\bibfnamefont {J.~S.}\ \bibnamefont
  {Miller}}\ and\ \bibinfo {editor} {\bibfnamefont {M.}~\bibnamefont
  {Drillon}}}\ (\bibinfo  {publisher} {Wiley-VCH Verlag GmbH \& Co},\ \bibinfo
  {address} {Weinheim, FRG},\ \bibinfo {year} {2002})\ pp.\ \bibinfo {pages}
  {345--466}\BibitemShut {NoStop}%
\bibitem [{\citenamefont {Chow}, \citenamefont {Chang},\ and\ \citenamefont
  {Willett}(1973)}]{Chow}%
  \BibitemOpen
  \bibfield  {author} {\bibinfo {author} {\bibfnamefont {C.}~\bibnamefont
  {Chow}}, \bibinfo {author} {\bibfnamefont {K.}~\bibnamefont {Chang}}, \ and\
  \bibinfo {author} {\bibfnamefont {R.~D.}\ \bibnamefont {Willett}},\ }\href
  {\doibase 10.1063/1.1680380} {\bibfield  {journal} {\bibinfo  {journal} {J.
  Chem. Phys.}\ }\textbf {\bibinfo {volume} {59}},\ \bibinfo {pages} {2629}
  (\bibinfo {year} {1973})}\BibitemShut {NoStop}%
\bibitem [{\citenamefont {Guo}\ \emph {et~al.}(2016)\citenamefont {Guo},
  \citenamefont {Watson}, \citenamefont {Hu}, \citenamefont {Sun},\ and\
  \citenamefont {Chan}}]{dmrg:nevpt2}%
  \BibitemOpen
  \bibfield  {author} {\bibinfo {author} {\bibfnamefont {S.}~\bibnamefont
  {Guo}}, \bibinfo {author} {\bibfnamefont {M.~A.}\ \bibnamefont {Watson}},
  \bibinfo {author} {\bibfnamefont {W.}~\bibnamefont {Hu}}, \bibinfo {author}
  {\bibfnamefont {Q.}~\bibnamefont {Sun}}, \ and\ \bibinfo {author}
  {\bibfnamefont {G.~K.-L.}\ \bibnamefont {Chan}},\ }\href {\doibase
  10.1021/acs.jctc.5b01225} {\bibfield  {journal} {\bibinfo  {journal} {J.
  Chem. Theory Comput.}\ }\textbf {\bibinfo {volume} {12}},\ \bibinfo {pages}
  {1583} (\bibinfo {year} {2016})}\BibitemShut {NoStop}%
\bibitem [{\citenamefont {Sharma}\ \emph {et~al.}(2014)\citenamefont {Sharma},
  \citenamefont {Sivalingam}, \citenamefont {Neese},\ and\ \citenamefont
  {Chan}}]{sharma:FeS}%
  \BibitemOpen
  \bibfield  {author} {\bibinfo {author} {\bibfnamefont {S.}~\bibnamefont
  {Sharma}}, \bibinfo {author} {\bibfnamefont {K.}~\bibnamefont {Sivalingam}},
  \bibinfo {author} {\bibfnamefont {F.}~\bibnamefont {Neese}}, \ and\ \bibinfo
  {author} {\bibfnamefont {G.~K.-L.}\ \bibnamefont {Chan}},\ }\href {\doibase
  10.1038/nchem.2041} {\bibfield  {journal} {\bibinfo  {journal} {Nature
  chemistry}\ }\textbf {\bibinfo {volume} {6}},\ \bibinfo {pages} {927}
  (\bibinfo {year} {2014})}\BibitemShut {NoStop}%
\bibitem [{\citenamefont {Neese}(2012)}]{orca}%
  \BibitemOpen
  \bibfield  {author} {\bibinfo {author} {\bibfnamefont {F.}~\bibnamefont
  {Neese}},\ }\href {\doibase 10.1002/wcms.81} {\bibfield  {journal} {\bibinfo
  {journal} {WIREs: Comput. Mol. Sci.}\ }\textbf {\bibinfo {volume} {2}},\
  \bibinfo {pages} {73} (\bibinfo {year} {2012})}\BibitemShut {NoStop}%
\bibitem [{\citenamefont {Bes}\ \emph {et~al.}(1999)\citenamefont {Bes},
  \citenamefont {Parisini}, \citenamefont {Inda}, \citenamefont {Saraiva},
  \citenamefont {Peleato},\ and\ \citenamefont {Sheldrick}}]{geom}%
  \BibitemOpen
  \bibfield  {author} {\bibinfo {author} {\bibfnamefont {M.~T.}\ \bibnamefont
  {Bes}}, \bibinfo {author} {\bibfnamefont {E.}~\bibnamefont {Parisini}},
  \bibinfo {author} {\bibfnamefont {L.~A.}\ \bibnamefont {Inda}}, \bibinfo
  {author} {\bibfnamefont {L.~M.}\ \bibnamefont {Saraiva}}, \bibinfo {author}
  {\bibfnamefont {M.~L.}\ \bibnamefont {Peleato}}, \ and\ \bibinfo {author}
  {\bibfnamefont {G.~M.}\ \bibnamefont {Sheldrick}},\ }\href {\doibase
  10.1016/S0969-2126(00)80054-4} {\bibfield  {journal} {\bibinfo  {journal}
  {Structure}\ }\textbf {\bibinfo {volume} {7}},\ \bibinfo {pages} {1201}
  (\bibinfo {year} {1999})}\BibitemShut {NoStop}%
\bibitem [{\citenamefont {Morales}\ \emph {et~al.}(1999)\citenamefont
  {Morales}, \citenamefont {Chron}, \citenamefont {Hudry-Clergeon},
  \citenamefont {Petillot}, \citenamefont {Norager}, \citenamefont {Medina},\
  and\ \citenamefont {Frey}}]{geom2}%
  \BibitemOpen
  \bibfield  {author} {\bibinfo {author} {\bibfnamefont {R.}~\bibnamefont
  {Morales}}, \bibinfo {author} {\bibfnamefont {M.}~\bibnamefont {Chron}},
  \bibinfo {author} {\bibfnamefont {G.}~\bibnamefont {Hudry-Clergeon}},
  \bibinfo {author} {\bibfnamefont {Y.}~\bibnamefont {Petillot}}, \bibinfo
  {author} {\bibfnamefont {S.}~\bibnamefont {Norager}}, \bibinfo {author}
  {\bibfnamefont {M.}~\bibnamefont {Medina}}, \ and\ \bibinfo {author}
  {\bibfnamefont {M.}~\bibnamefont {Frey}},\ }\href {\doibase
  10.1021/bi991578s} {\bibfield  {journal} {\bibinfo  {journal} {Biochemistry}\
  }\textbf {\bibinfo {volume} {38}},\ \bibinfo {pages} {15764} (\bibinfo {year}
  {1999})}\BibitemShut {NoStop}%
\bibitem [{\citenamefont {Peng}\ \emph {et~al.}(2013)\citenamefont {Peng},
  \citenamefont {Middendorf}, \citenamefont {Weigend},\ and\ \citenamefont
  {Reiher}}]{x2c}%
  \BibitemOpen
  \bibfield  {author} {\bibinfo {author} {\bibfnamefont {D.}~\bibnamefont
  {Peng}}, \bibinfo {author} {\bibfnamefont {N.}~\bibnamefont {Middendorf}},
  \bibinfo {author} {\bibfnamefont {F.}~\bibnamefont {Weigend}}, \ and\
  \bibinfo {author} {\bibfnamefont {M.}~\bibnamefont {Reiher}},\ }\href
  {\doibase 10.1063/1.4803693} {\bibfield  {journal} {\bibinfo  {journal} {J.
  Chem. Phys.}\ }\textbf {\bibinfo {volume} {138}},\ \bibinfo {pages} {184105}
  (\bibinfo {year} {2013})}\BibitemShut {NoStop}%
\bibitem [{\citenamefont {Cheng}\ \emph {et~al.}(1994)\citenamefont {Cheng},
  \citenamefont {Xia}, \citenamefont {Reed},\ and\ \citenamefont
  {Markley}}]{Cheng}%
  \BibitemOpen
  \bibfield  {author} {\bibinfo {author} {\bibfnamefont {H.}~\bibnamefont
  {Cheng}}, \bibinfo {author} {\bibfnamefont {B.}~\bibnamefont {Xia}}, \bibinfo
  {author} {\bibfnamefont {G.~H.}\ \bibnamefont {Reed}}, \ and\ \bibinfo
  {author} {\bibfnamefont {J.~L.}\ \bibnamefont {Markley}},\ }\href {\doibase
  10.1021/bi00177a003} {\bibfield  {journal} {\bibinfo  {journal}
  {Biochemistry}\ }\textbf {\bibinfo {volume} {33}},\ \bibinfo {pages} {3155}
  (\bibinfo {year} {1994})}\BibitemShut {NoStop}%
\bibitem [{\citenamefont {Fritz}\ \emph {et~al.}(1971)\citenamefont {Fritz},
  \citenamefont {Anderson}, \citenamefont {Fee}, \citenamefont {Palmer},
  \citenamefont {Sands}, \citenamefont {Tsibris}, \citenamefont {Gunsalus},
  \citenamefont {Orme-Johnson},\ and\ \citenamefont {Beinert}}]{fritz}%
  \BibitemOpen
  \bibfield  {author} {\bibinfo {author} {\bibfnamefont {J.}~\bibnamefont
  {Fritz}}, \bibinfo {author} {\bibfnamefont {R.}~\bibnamefont {Anderson}},
  \bibinfo {author} {\bibfnamefont {J.}~\bibnamefont {Fee}}, \bibinfo {author}
  {\bibfnamefont {G.}~\bibnamefont {Palmer}}, \bibinfo {author} {\bibfnamefont
  {R.~H.}\ \bibnamefont {Sands}}, \bibinfo {author} {\bibfnamefont {J.~C.~M.}\
  \bibnamefont {Tsibris}}, \bibinfo {author} {\bibfnamefont {J.~C.}\
  \bibnamefont {Gunsalus}}, \bibinfo {author} {\bibfnamefont {W.~H.}\
  \bibnamefont {Orme-Johnson}}, \ and\ \bibinfo {author} {\bibfnamefont
  {H.}~\bibnamefont {Beinert}},\ }\href {\doibase 10.1016/0005-2728(71)90239-8}
  {\bibfield  {journal} {\bibinfo  {journal} {Biochim. Biophys. Acta}\ }\textbf
  {\bibinfo {volume} {253}},\ \bibinfo {pages} {110} (\bibinfo {year}
  {1971})}\BibitemShut {NoStop}%
\bibitem [{\citenamefont {Sands}\ and\ \citenamefont {Dunham}(1975)}]{Fe2S2_g}%
  \BibitemOpen
  \bibfield  {author} {\bibinfo {author} {\bibfnamefont {R.~H.}\ \bibnamefont
  {Sands}}\ and\ \bibinfo {author} {\bibfnamefont {W.~R.}\ \bibnamefont
  {Dunham}},\ }\href {\doibase 10.1017/S0033583500001517} {\bibfield  {journal}
  {\bibinfo  {journal} {Quart. Rev. Biophys.}\ }\textbf {\bibinfo {volume}
  {4}},\ \bibinfo {pages} {443} (\bibinfo {year} {1975})}\BibitemShut {NoStop}%
\bibitem [{\citenamefont {Gayda}\ \emph {et~al.}(1981)\citenamefont {Gayda},
  \citenamefont {Bertrand}, \citenamefont {More},\ and\ \citenamefont
  {Cammack}}]{gayda3}%
  \BibitemOpen
  \bibfield  {author} {\bibinfo {author} {\bibfnamefont {J.-P.}\ \bibnamefont
  {Gayda}}, \bibinfo {author} {\bibfnamefont {P.}~\bibnamefont {Bertrand}},
  \bibinfo {author} {\bibfnamefont {C.}~\bibnamefont {More}}, \ and\ \bibinfo
  {author} {\bibfnamefont {R.}~\bibnamefont {Cammack}},\ }\href {\doibase
  10.1016/S0300-9084(82)80271-X} {\bibfield  {journal} {\bibinfo  {journal}
  {Biochimie}\ }\textbf {\bibinfo {volume} {63}},\ \bibinfo {pages} {847}
  (\bibinfo {year} {1981})}\BibitemShut {NoStop}%
\bibitem [{\citenamefont {Malkin}\ and\ \citenamefont
  {Bearden}(1971)}]{Malkin}%
  \BibitemOpen
  \bibfield  {author} {\bibinfo {author} {\bibfnamefont {R.}~\bibnamefont
  {Malkin}}\ and\ \bibinfo {author} {\bibfnamefont {A.~J.}\ \bibnamefont
  {Bearden}},\ }\href@noop {} {\bibfield  {journal} {\bibinfo  {journal} {Proc.
  Natl. Acad. Sci.}\ }\textbf {\bibinfo {volume} {68}},\ \bibinfo {pages} {16}
  (\bibinfo {year} {1971})}\BibitemShut {NoStop}%
\bibitem [{\citenamefont {Gibson}\ \emph {et~al.}(1966)\citenamefont {Gibson},
  \citenamefont {Hall}, \citenamefont {Thornley},\ and\ \citenamefont
  {Whatley}}]{gibson}%
  \BibitemOpen
  \bibfield  {author} {\bibinfo {author} {\bibfnamefont {J.~F.}\ \bibnamefont
  {Gibson}}, \bibinfo {author} {\bibfnamefont {D.~O.}\ \bibnamefont {Hall}},
  \bibinfo {author} {\bibfnamefont {J.~H.~M.}\ \bibnamefont {Thornley}}, \ and\
  \bibinfo {author} {\bibfnamefont {F.~R.}\ \bibnamefont {Whatley}},\ }\href
  {\doibase 10.1073/pnas.56.3.987} {\bibfield  {journal} {\bibinfo  {journal}
  {Proc. Natl. Acad. Sci.}\ }\textbf {\bibinfo {volume} {56}},\ \bibinfo
  {pages} {987} (\bibinfo {year} {1966})}\BibitemShut {NoStop}%
\bibitem [{\citenamefont {Mukai}\ \emph {et~al.}(1973)\citenamefont {Mukai},
  \citenamefont {Kimura}, \citenamefont {Helbert},\ and\ \citenamefont
  {Kevan}}]{Mukai}%
  \BibitemOpen
  \bibfield  {author} {\bibinfo {author} {\bibfnamefont {K.}~\bibnamefont
  {Mukai}}, \bibinfo {author} {\bibfnamefont {T.}~\bibnamefont {Kimura}},
  \bibinfo {author} {\bibfnamefont {J.}~\bibnamefont {Helbert}}, \ and\
  \bibinfo {author} {\bibfnamefont {L.}~\bibnamefont {Kevan}},\ }\href
  {\doibase 10.1016/0005-2795(73)90072-X} {\bibfield  {journal} {\bibinfo
  {journal} {Biochim. Biophys. Acta}\ }\textbf {\bibinfo {volume} {295}},\
  \bibinfo {pages} {49} (\bibinfo {year} {1973})}\BibitemShut {NoStop}%
\bibitem [{\citenamefont {Palmer}, \citenamefont {Sands},\ and\ \citenamefont
  {Mortenson}(1966)}]{Palmer_g}%
  \BibitemOpen
  \bibfield  {author} {\bibinfo {author} {\bibfnamefont {G.}~\bibnamefont
  {Palmer}}, \bibinfo {author} {\bibfnamefont {R.~H.}\ \bibnamefont {Sands}}, \
  and\ \bibinfo {author} {\bibfnamefont {L.~E.}\ \bibnamefont {Mortenson}},\
  }\href {\doibase 10.1016/0006-291X(66)90733-9} {\bibfield  {journal}
  {\bibinfo  {journal} {Biochim. Biophys. Res. Commun}\ }\textbf {\bibinfo
  {volume} {23}},\ \bibinfo {pages} {357} (\bibinfo {year} {1966})}\BibitemShut
  {NoStop}%
\bibitem [{\citenamefont {Pantazis}\ \emph {et~al.}(2012)\citenamefont
  {Pantazis}, \citenamefont {Ames}, \citenamefont {Cox}, \citenamefont
  {Lubitz},\ and\ \citenamefont {Neese}}]{oec}%
  \BibitemOpen
  \bibfield  {author} {\bibinfo {author} {\bibfnamefont {D.~A.}\ \bibnamefont
  {Pantazis}}, \bibinfo {author} {\bibfnamefont {W.}~\bibnamefont {Ames}},
  \bibinfo {author} {\bibfnamefont {N.}~\bibnamefont {Cox}}, \bibinfo {author}
  {\bibfnamefont {W.}~\bibnamefont {Lubitz}}, \ and\ \bibinfo {author}
  {\bibfnamefont {F.}~\bibnamefont {Neese}},\ }\href {\doibase DOI:
  10.1002/anie.201204705} {\bibfield  {journal} {\bibinfo  {journal} {Angew.
  Chem. Int. Ed.}\ }\textbf {\bibinfo {volume} {51}},\ \bibinfo {pages} {9935}
  (\bibinfo {year} {2012})}\BibitemShut {NoStop}%
\bibitem [{\citenamefont {Kurashige}, \citenamefont {Chan},\ and\ \citenamefont
  {Yanai}(2013)}]{kurashige2013entangled}%
  \BibitemOpen
  \bibfield  {author} {\bibinfo {author} {\bibfnamefont {Y.}~\bibnamefont
  {Kurashige}}, \bibinfo {author} {\bibfnamefont {G.~K.-L.}\ \bibnamefont
  {Chan}}, \ and\ \bibinfo {author} {\bibfnamefont {T.}~\bibnamefont {Yanai}},\
  }\href {\doibase 10.1038/nchem.1677} {\bibfield  {journal} {\bibinfo
  {journal} {Nature chemistry}\ }\textbf {\bibinfo {volume} {5}},\ \bibinfo
  {pages} {660} (\bibinfo {year} {2013})}\BibitemShut {NoStop}%
\bibitem [{\citenamefont {Dismukes}\ and\ \citenamefont
  {Sidererse}(1981)}]{oec2}%
  \BibitemOpen
  \bibfield  {author} {\bibinfo {author} {\bibfnamefont {G.~C.}\ \bibnamefont
  {Dismukes}}\ and\ \bibinfo {author} {\bibfnamefont {Y.}~\bibnamefont
  {Sidererse}},\ }\href {\doibase 10.1073/pnas.78.1.274} {\bibfield  {journal}
  {\bibinfo  {journal} {Proc. Natl. Acad. Sci. USA}\ }\textbf {\bibinfo
  {volume} {78}},\ \bibinfo {pages} {274} (\bibinfo {year} {1981})}\BibitemShut
  {NoStop}%
\bibitem [{\citenamefont {Hansson}\ and\ \citenamefont {Andr{\'
  e}asson}(1982)}]{oec3}%
  \BibitemOpen
  \bibfield  {author} {\bibinfo {author} {\bibfnamefont {{\"O}.}~\bibnamefont
  {Hansson}}\ and\ \bibinfo {author} {\bibfnamefont {L.-E.}\ \bibnamefont
  {Andr{\' e}asson}},\ }\href {\doibase 10.1016/0005-2728(82)90296-1}
  {\bibfield  {journal} {\bibinfo  {journal} {Biochim. Biophys. Acta}\ }\textbf
  {\bibinfo {volume} {679}},\ \bibinfo {pages} {261} (\bibinfo {year}
  {1982})}\BibitemShut {NoStop}%
\bibitem [{\citenamefont {Vinyard}\ \emph {et~al.}(2017)\citenamefont
  {Vinyard}, \citenamefont {Khan}, \citenamefont {Askerka}, \citenamefont
  {Batista},\ and\ \citenamefont {Brudvig}}]{oec4}%
  \BibitemOpen
  \bibfield  {author} {\bibinfo {author} {\bibfnamefont {D.~J.}\ \bibnamefont
  {Vinyard}}, \bibinfo {author} {\bibfnamefont {S.}~\bibnamefont {Khan}},
  \bibinfo {author} {\bibfnamefont {M.}~\bibnamefont {Askerka}}, \bibinfo
  {author} {\bibfnamefont {V.~S.}\ \bibnamefont {Batista}}, \ and\ \bibinfo
  {author} {\bibfnamefont {G.~W.}\ \bibnamefont {Brudvig}},\ }\href {\doibase
  10.1021/acs.jpcb.7b00110} {\bibfield  {journal} {\bibinfo  {journal} {J.
  Phys. Chem. B}\ }\textbf {\bibinfo {volume} {121}},\ \bibinfo {pages} {1020}
  (\bibinfo {year} {2017})}\BibitemShut {NoStop}%
\end{thebibliography}%

\end{document}